\documentclass{emulateapj-rtx4}

\usepackage{natbib}
\usepackage{lscape}

\bibpunct{(}{)}{;}{a}{}{,}

\newcommand{\ee}[1]{\mbox{${} \times 10^{#1}$}}
\newcommand{\eten}[1]{\mbox{$10^{#1}$}}

\newcommand\cmn{\mbox{cm$^{-2}$}}

\newcommand{\lsun}{\mbox{L$_\odot$}}
\newcommand{\msun}{\mbox{M$_\odot$}}

\newcommand{\lint}{\mbox{$L_{int}$}} 

\newcommand{\tbnm}{\tablenotemark}
\newcommand{\tbnt}{\tablenotetext}

\newcommand{\water}{H$_2$O}

\newcommand{\cooo}{C$^{18}$O}

\newcommand{\cotwo}{\mbox{CO$_2$}}
\newcommand{\jj}[2]{\mbox{$J = #1\rightarrow#2$}}

\begin{document}

\title {CO$_2$ Ice toward Low-luminosity, Embedded Protostars:
Evidence for Episodic Mass Accretion via Chemical History
}

\author{
Hyo Jeong Kim\altaffilmark{1,2},
Neal J. Evans II\altaffilmark{1},
Michael M. Dunham\altaffilmark{3},
Jeong-Eun Lee\altaffilmark{4},
Klaus M. Pontoppidan\altaffilmark{5}
}

\altaffiltext{1}{Department of Astronomy, The University of Texas at
Austin, 2515 Speedway, Stop C1400 Austin, TX 78712-1205, USA}

\altaffiltext{2}{E-mail: hyojeong@astro.as.utexas.edu}

\altaffiltext{3}{Department of Astronomy, Yale University, P.O. Box
  208101, New Haven, CT 06520, USA}

\altaffiltext{4}{Department of Astronomy and Space Science, Kyung Hee
  University, Yongin-si, Gyeonggi-do 446-701, Korea.}

\altaffiltext{5}{Space Telescope Science Institute, Baltimore, MD
  21218, USA}

\begin{abstract}

We present \emph{Spitzer} IRS spectroscopy of \cotwo\ ice bending
mode spectra at 15.2 \micron\ toward 19 young stellar objects with
luminosity lower than 1 L$_{\odot}$ (3 with luminosity lower than
0.1 L$_{\odot}$). Ice on dust grain surfaces can encode the history
of heating because pure \cotwo\ ice forms only at elevated
temperature, $T > 20$ K, and thus around protostars of higher
luminosity. Current internal luminosities of YSOs with $L < 1$
\lsun\ do not provide the conditions needed to produce pure \cotwo\
ice at radii where typical envelopes begin. The presence of
detectable amounts of pure \cotwo\ ice would signify a higher past
luminosity. Many of the spectra require a contribution from a pure,
crystalline \cotwo\ component, traced by the presence of a
characteristic band splitting in the 15.2 \micron\ bending mode.
About half of the sources (9 out of 19) in the low luminosity
sample have evidence for pure \cotwo\ ice, and six of these have
significant double-peaked features, which are very strong evidence
of pure \cotwo\ ice. The presence of the pure \cotwo\ ice component
indicates that the dust temperature, and hence luminosity of the
central star/accretion disk system, must have been higher in the
past. An episodic accretion scenario, in which mixed CO-\cotwo\ ice
is converted to pure \cotwo\ ice during each high luminosity phase,
explains the presence of pure \cotwo\ ice, the total amount of
\cotwo\ ice, and the observed residual \cooo\ gas.

\end{abstract}



\section{Introduction}\label{intro}

While the standard star formation model with constant accretion rate 
\citep{1977ApJ...214..488S, 1987ARA&A..25...23S}
predicts protostars to have a luminosity higher than 1.6 \lsun\ 
most of the time, young stellar objects (YSOs)
with luminosity lower than 1.6 \lsun\ have long been known 
\citep{1990AJ.....99..869K,
1994ApJ...434..614G, 1996ARA&A..34..207H}. 
The \emph{Spitzer} Legacy
Project, From Molecular Cores to Planet Forming Disks (c2d;
\citealt{2003PASP..115..965E}) found that 59\% of the 112 embedded
protostars have luminosity lower than 1.6 L$_{\odot}$
\citep{2009ApJS..181..321E}. In fact, \emph{Spitzer} found a
substantial number of sources with the internal luminosity even
below 0.1 \lsun\ (\citealt{2004ApJS..154..396Y,
  2008ApJS..179..249D}), which have been called Very Low Luminosity
Objects (VeLLOs; \citealt{2007prpl.conf...17D}). The internal
luminosity (\lint) of an embedded protostar is the luminosity of the central
source, excluding luminosity from external heating. Recently, 
an even lower luminosity protostar ($\lint < 0.03$
\lsun) was discovered as a binary companion to IRAM04191+1522 IRS
\citep{2012ApJ...747L..43C}. One explanation for such low luminosities
is that mass accretion is not a constant process
\citep{1990AJ.....99..869K, 2009ApJ...692..973E,
  2010ApJ...710..470D}. The low luminosity sources may be going
through a low mass accretion stage between accretion bursts, explaining 
their currently low luminosities. If mass accretion is episodic,
sources with low luminosity may have had much higher accretion rates
in the past, and thus had higher luminosity. Imprints of the high
luminosity stage in low luminosity YSOs would support the idea of
episodic mass accretion.

Evidence of past periods of higher luminosity
 can be found in molecular spectra of the gas and ice phases
of low luminosity sources 
(\citealt{2007JKAS...40...85L, 2012ApJ...754L..18V}). 
In particular, \citet{2008ApJ...678.1005P} analyzed the
general shape of the 15.2 \micron~CO$_2$ ice bending mode spectrum
toward 50 embedded young stars. The 15.2 \micron~CO$_2$ spectrum can
be decomposed into multiple components, including a pure CO$_2$ ice
component. The pure CO$_2$ ice can form by two processes. One is
CO$_2$  segregation out of a CO$_2$-H$_2$O mixture. The other is a
distillation process, in  which CO evaporates from a CO$_2$-CO
mixture, leaving pure CO$_2$ behind. The former process occurs at a
high temperature (50 \-- 80 K) and the latter occurs at a lower
temperature (20 \-- 30 K). Both pure CO$_2$ formation processes are
irreversible \citep{1983A&AS...51..389H}, since the bond between
pure CO$_2$ ices makes it the most stable phase. Once the pure CO$_2$
ice has formed, it will not disappear unless it evaporates. Thus the
existence of pure \cotwo\ ice provides a ``chemical memory" of warmer
conditions in the past.

The total amount of CO$_2$ ice, regardless of form, is another
indicator of the temperature history. A number of studies of
absorption against background stars have shown that \cotwo\ ice 
must form in the prestellar phase 
(\citealt{2005ApJ...627L..33B, 2005ApJ...635L.145K, 2009ApJ...695...94W,
 2011ApJ...731....9C}).
The shape of the 15.2 \micron\ feature toward background stars
varies little from cloud to cloud 
\citep{2009ApJ...695...94W},
with most ($\sim 85$ to 90\%) of the \cotwo\ in the \cotwo-\water\ mixture
and no evidence for a pure \cotwo\ component, despite variations in the 
ratio of CO to \water\ ice from cloud to cloud \citep{2009ApJ...695...94W}.
These ices arise from reactions among H, O, and CO on grain surfaces
(e.g., \citealt{2011ApJ...731....9C}).


Observations of the \cotwo\ ice feature toward YSOs tend to show 
higher values of $N(\cotwo)$ than the background stars
\citep{2008ApJ...678.1005P}, 
even for background stars with roughly the same extinction (see Fig. 5
in \citealt{2011ApJ...730..124C}).
The ratio of \cotwo\ ice to \water\ ice in spectra toward YSOs is also
higher and more variable than toward background stars while the ratio
of CO to \water\ ice is lower for the YSOs, indicative of further
processing occuring during the formation of the YSOs 
\citep{2011ApJ...730..124C}.

The total amount of CO$_2$ ice formed during the YSO phase
depends on the accretion scenario.  The episodic accretion 
scenario gives multiple long periods of low luminosity
\citep{2010ApJ...710..470D} instead of the very short period of low
luminosity that the continuous accretion model predicts
\citep{1977ApJ...214..488S, 2005ApJ...627..293Y}. As a result,
episodic accretion provides more time for CO to freeze-out
and form \cotwo\ ice.  Then, the CO ice can evaporate during
episodes of higher accretion and luminosity, leaving the 
\cotwo\ ice behind.

The gas phase CO is the final indicator that we consider. Laboratory
experiments show that the binding energies of CO and N$_2$ onto dust
grain surfaces are similar \citep{2005ApJ...621L..33O,
2006A&A...449.1297B}. After including the updated binding energy,
chemo-dynamical models \citep{2004ApJ...617..360L} predict more CO,
as measured by rare isotopes like \cooo, than is observed (e.g.,
\citealt{2009ApJ...705.1160C, 2011ApJ...729...84K}).
Tying up some of the carbon in \cotwo\ ice could help to match
the observations \citep{2012ApJ...754L..18V}.

To explain the observed gas phase molecular lines and the total
CO$_2$ ice absorption observed toward one specific low luminosity
source, CB130-1-IRS1, \citet{2011ApJ...729...84K} added some simple
ice reactions to an episodic accretion chemical model
(\citealt{2004ApJ...617..360L, 2007JKAS...40...85L}). During low
luminosity periods, the gas phase CO freezes into CO ice, and some
CO ice turns into CO$_2$ ice, creating a CO-\cotwo\ ice mixture;
then, during high luminosity phases, the CO evaporates, leaving
behind \cotwo\ ice, which evaporates only at a still higher
temperature ($T \geq 20$ K), depending on the structure of the ice
\citep{2008ApJ...678.1005P}. This scenario, developed by
\citet{2011ApJ...729...84K} and discussed further by 
\citet{2012ApJ...754L..18V},
could explain the current
low luminosity, the total amount of CO$_2$ ice, and the strength of
the gas phase molecule emission toward CB130-1-IRS1.

In this paper, we test the idea further with higher resolution
Infrared Spectrograph (IRS; \citealt{2004ApJS..154...18H})
spectra for an enlarged source sample and complementary spectroscopy
of gas-phase \cooo\ at 1.3 mm. We obtained \emph{Spitzer} IRS,
Short-High (SH) mode spectroscopy toward 19 young stellar objects
(YSOs) with luminosity lower than 1 L$_{\odot}$, 4 of them with
luminosity lower than 0.1 L$_{\odot}$. The envelopes around YSOs
with luminosity lower than 1 L$_{\odot}$ have temperatures lower
than the CO evaporation temperature. The higher resolution spectra
allow decomposition of the 15.2 \micron~ CO$_2$ ice bending mode
spectrum into 5 different components, including a pure \cotwo\
component, using the method of \citet{2008ApJ...678.1005P}. In
Section \ref{sec:sources}, we explain the sample selection in the
current study. We describe the observations in Section
\ref{SEC_Observations}. In Section \ref{sec:analysis}, we describe
the analysis of the \emph{Spitzer} IRS spectrum to get the optical
depth and the column density. In Section \ref{SEC_RESULTS} we
present results of the decomposition of ice components. In Section
\ref{sec:Discuss}, we present the setup and results of the chemical
models for different accretion scenarios. In Section
\ref{sec:conclusion}, we summarize our findings.

\section{Sources}\label{sec:sources}
The sources were selected from the list of low luminosity objects
compiled by \citet{2008ApJS..179..249D}. Table~\ref{table-source}
includes the source identification numbers and internal luminosities
from that reference, the \emph{Spitzer} source name, the cores or
clouds with which a source is associated, \emph{Spitzer} positions,
and the Program ID (PID) of the \emph{Spitzer} IRS observation. The
internal luminosities are determined by using the  correlation between
70 $\micron$ flux and the internal luminosity of protostars
\citep{2008ApJS..179..249D}. The sources are Class 0/I objects with
luminosities in  the range between 0.08 \-- 0.69 \lsun. We also use
sources from \citet{2008ApJ...678.1005P}, most of which have
luminosities between 1 and 10 \lsun, but some are massive sources with
luminosities up to 10$^5$ \lsun. With the two samples, we can study
objects with luminosities from 0.1 \lsun~to 10$^5$ \lsun.

\section{Observations}\label{SEC_Observations}
The \emph{Spitzer} IRS observations of 17 sources were obtained
between November 2008 and May 2009 (PID 50295; PI Michael Dunham)
with their background image. Spectra of two more sources, source
number 003 and 038, were obtained from \emph{Spitzer} archival data,
observed in 2005, October and November respectively (PID 20604; PI
Adwin Boogert). 
Background images were not observed for these two sources. 
The observing mode was SH, and the resolving
power ($\lambda / \Delta \lambda$) was 600. The data were
reduced following the IRS pipeline version S18.7 which produces
Basic Calibrated Data (BCD) files. Then we used the bad pixel masks
to clean the masked pixels in a set of data. We used the Caltech
High-res IRS pipeline (CHIP) to reduce the 17 observed data with
background exposure. CHIP is an IDL reduction package for reducing
high signal to noise SH \emph{Spitzer} IRS spectra 
\citep{2010ApJ...720..887P}.  CHIP produces better signal to noise level 
but requires the dedicated background exposures. 
For the two sources without background images, we used SMART
\citep{2004PASP..116..975H, 2010PASP..122..231L} to do the
background subtraction from their own image and to extract spectra
from the BCD files. Then we trimmed the end of each order of a
spectrum. 
All 19 sources have clear detections of the CO$_2$ ice
bending mode absorption spectrum. The overlapping regions between
orders match well.

The gas-phase molecular line observations were obtained at the CSO
\footnote{The Caltech Submillimeter Observatory was supported by the
NSF.} in 2010 July. The \cooo\ (\jj{2}1) lines were obtained with
the fast fourier transform spectrometer (FFTS) having 8192 channels
with a 500 MHz total bandwidth. We used the 230 GHz heterodyne receiver in
position switching mode. The main beam efficiency was 0.8,  and the
velocity resolution was 0.17 km/s.

\section{Analysis}\label{sec:analysis}
\subsection{Reduction of the IRS Spectra}\label{sec:continuum}

\begin{figure}[t]
\includegraphics{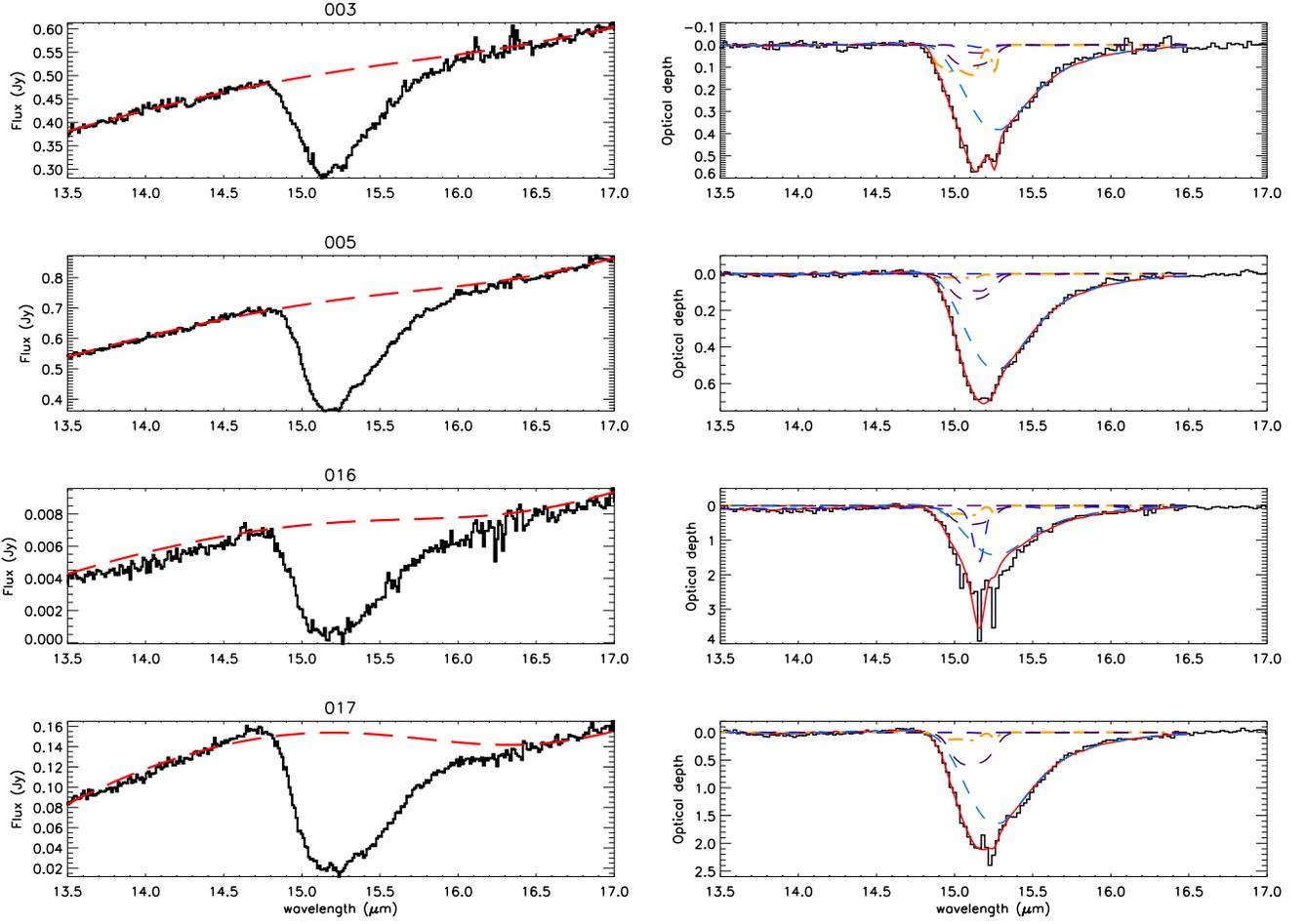}
\caption{\label{figure1:spectrum} Left panels : observed flux
  (black solid line) and best fit continuum (red dashed line). Right
  panels : optical depth (black solid line), sum of all the ice
  components (red solid line, best fit model), pure CO$_2$ (yellow
  dash-dot), H$_2$O-rich CO$_2$ ice component (blue dashed), CO-CO$_2$
  mixtures (purple dashed).}
\end{figure}

The 15.2 \micron~CO$_2$ bending mode spectrum is located between the
broad 9.7 \micron, and 18 \-- 20 \micron~silicate features, and near
the 12 \micron\ H$_2$O band. To remove the baseline, we constructed
a third-order polynomial in the range of 13.3 $\--$ 14.7 \micron\
and 16.2 $\--$ 19.3 \micron,. and added a gaussian feature centered
at 608 cm$^{-1}$ with FWHM of 73 cm$^{-1}$ for the blue wing of the
18 \-- 20 \micron~silicate feature \citep{2008ApJ...678.1005P}. The
15.2 \micron~CO$_2$ bending mode spectra are plotted in
Fig.~\ref{figure1:spectrum} \-- Fig.~\ref{Ophiuchius:spectrum}. The
best fit baselines are given as red dashed lines in the left panels
of Fig.~\ref{figure1:spectrum} \-- Fig.~\ref{Ophiuchius:spectrum}.
After the continuum subtraction, we calculated the optical depth of
each source by $\tau = - ln \left(\frac{I}{I_0} \right )$, where
$I_0$ is the continuum intensity. Optical depths are plotted in the
right panels of Fig. ~\ref{figure1:spectrum} \--
Fig.~\ref{Ophiuchius:spectrum}.

\subsection{Decomposition into Ice Components}\label{sec:labodat}

We used a number of laboratory spectra of CO$_2$ ices to analyze the
15.2 \micron\ CO$_2$ bending mode. The available laboratory spectra are
CO$_2$:H$_2$O mixtures, CO$_2$:CO mixtures
\citep{1997A&A...328..649E}, and pure CO$_2$
\citep{2006A&A...451..723V}. A minimum of five unique components has
been required to fit all previously observed 15.2 \micron\ CO$_2$
spectra \citep{2008ApJ...678.1005P}. They are a dilute CO$_2$
(CO$_2$:CO = 4:100), an H$_2$O-rich component (CO$_2$:H$_2$O =
14:100), a mixture of CO and CO$_2$ (CO$_2$:CO = 26:100 and
CO$_2$:CO =70:100), a pure CO$_2$ component, and a shoulder
component. We did not include the shoulder component, since the
shoulder component appears only in massive young stars, but we use
both CO$_2$:CO = 26:100 and CO$_2$:CO =70:100 mixtures as separate
components. Thus, the number of CO$_2$ ice components is still 5 in
the current study. We use the laboratory data with the grain shape
of the continuous distribution of ellipsoids (CDE) to convert
optical constants to  opacities. This is a method to model
irregularly shaped grains, successfully used in
\citet{2008ApJ...678.1005P}. The laboratory ices have been obtained
at temperatures between 10 K and  130 K. The laboratory data are not
strongly dependent on the temperature. For example, the strength of
the pure CO$_2$ ice component decreases by 1\% when the temperature
increases from 10 K to 50 K. When temperature increases to 80 K, the
strength of the component decreases by 9\%. The envelopes of low
luminosity sources have temperatures lower than 20 K. We use the
laboratory ices with a temperature at 10 K, except that the pure
CO$_2$ component is at 15 K, for which the laboratory data are
available.

The laboratory data for the H$_2$O-rich CO$_2$ ice component lies on
top of a broad 12 \micron\ H$_2$O ice feature. So we fitted a
baseline with a third order polynomial, and then subtracted the
baseline from the laboratory data. The resulting laboratory data
used to decompose \cotwo\ ice features are plotted in
Fig.~\ref{labo:dat}. The dilute CO$_2$ component is located at the
red part of 15.2 \micron~bending mode. The CO$_2$:CO = 70:100
component is located at the blue part of 15.2 \micron~bending mode.
Only the pure CO$_2$ ice component shows a double peaked feature.

We varied only the strength of each component to fit the observed optical
depth of the CO$_2$ feature. 
Then we used a $\chi ^2$ minimization technique
to obtain the best fit. We have 5 free parameters, the strengths for
each of the 5 different ice components. The reduced $\chi^2$ of the
best fit is listed in Table~\ref{table-columnD}. The best fit
laboratory components are plotted separately and in total in the
right panels of Fig.~\ref{figure1:spectrum} \--
Fig.~\ref{Ophiuchius:spectrum}. In all sources, the water-rich
CO$_2$ ice dominates the shape and the strength of the spectra. The
red wing part of the spectrum is determined by the water-rich
component. The peak strength is determined by the sum of all the
components. The double peaked features, unique to the pure \cotwo\
ice, generally agree in wavelength with the laboratory data.
We did the component analysis with and without the pure \cotwo\ 
ice component included in the fit.

\subsection{Column Densities of Ices}\label{sec:columnN}
To obtain a column density from the optical depth, we integrated the
optical depth over wave number, and then divided it by the band strength,
A:
\begin{equation}
N (cm ^{-2})=\int \tau (\nu ) d\nu / A .
\end{equation}
We used the band strength from \citet{1995A&A...296..810G}. For the
pure CO$_2$ component, the CO$_2$:CO = 4:100 component, and the
CO$_2$:CO = 26:100 component, $A =  1.1 \times 10^{-17}$ cm
molecule$^{-1}$. For CO$_2$:CO =70:100, $A =  0.97 \times 10^{-17}$ cm
molecule$^{-1}$, and the component with CO$_2$:H$_2$O = 14:100 has $A
=  1.57 \times 10^{-17}$ cm molecule$^{-1}$. The column densities for
each component are listed in Table ~\ref{table-columnD}.

\subsection{Column Density of Gas Phase CO}\label{sec:columngas}

The CSO data were reduced with CLASS
\footnote{http://www.iram.fr/IRAMFR/GILDAS}. The molecular line
results are summarized in Table \ref{table:molecularline}. The
column density of \cooo\ is obtained from

\begin{equation}
N_{C^{18}O}=\frac{3 k Q ~exp(E_J /kT_{ex})}{8 \pi^3 \nu \mu^2 J} \int {T_R} ~dv,
\end{equation}
where $E_J = hBJ(J+1)$, B is the rotational constant, $T_{ex}$ is the
excitation temperature, $\mu$ is the dipole moment, and $Q$ is the
partition function. The parameters B and $\mu$ for \cooo\ come from
Table 3 of \citet{2003ApJ...583..789L}. We assumed $T_{ex} = 10$ K
to derive the column density.

\section{Results}\label{SEC_RESULTS}

Double peaked features are clearly seen for six sources (30\%),
as indicated in Table \ref{table-columnD}.
As Fig.~\ref{labo:dat} indicates, only the pure CO$_2$ ice component
has a double peak at wavelengths 15.14 \micron\ and 15.26 \micron.
All the sources with obvious double-peaked features have fits for
column density of the pure \cotwo\ component that are more than 10
$\sigma$, and the value of $\chi^2$ improved by more than 20\% when the pure
\cotwo\ ice component was included. There are three other sources without
obvious double-peaked features that satisfy the same criteria for 
signal-to-noise and improvement in $\chi^2$, also marked in Table
\ref{table-columnD}. 
We also consider these to be detections, for a total of 9 detections
out of 19 sources (47\%).
In fact, all but 3 sources have best fit values of \cotwo\ ice column
densities with signal-to-noise greater than 10, but we require the improvement
in $\chi^2$ because uncertainties in a multi-component fit may be unreliable.

We detected the pure CO$_2$ ice component with a significant double
peak even from the source with $\lint = 0.08$ \lsun\ (Source number:
003). \citet{2008ApJ...678.1005P} found the pure CO$_2$ ice
in 40\% of the total 56 sources with clear detections in sources
with luminosity higher than 1 \lsun. The current result shows that
the presence of pure CO$_2$ ice is at least as common in low
luminosity protostars as it is in high luminosity protostars.
This is all the more remarkable because absorption in the
surrounding molecular cloud from unprocessed \cotwo\ ice 
\citep{2009ApJ...695...94W}
will make the pure \cotwo\ ice signal harder to see.

CO evaporates from CO$_2$-CO ice mixture at a temperature 20 \-- 30
K. We take the minimum required temperature for pure CO$_2$ ice
formation as 20 K. It is beyond the scope of this paper to explore
the parameter space of the structures of the protostellar envelopes,
but through model studies, the envelope inner radii of embedded
protostars are $\geq 100$ AU. (Serpens 1: 600 AU
\citealt{2009ApJ...707..103E}, IRAM 04191: 140 AU
\citealt{2006ApJ...651..945D}, CB130-1: 350 AU
\citealt{2011ApJ...729...84K}, L673-7: 140 AU
\citealt{2010ApJ...721..995D}). Many of these
individual studies used simple 1-D or 2-D geometry to approximate
the complicated transition between the envelope and disk, and the
envelope inner radius is a way of parameterizing this transition
with a single value.  To match observed SEDs, one must have an inner
cut-off to a spherical density profile to avoid having extremely
high density material, and thus optical depth. Though 100 AU is not
an exact value representing the inner radii of all envelopes, it is
a minimum value compared to values found in many models.

Using DUSTY \citep{1999astro.ph.10475I}, we calculated a dust
temperature around sources with luminosities of 0.7 \lsun\ (the
highest current luminosity in our sample) and 150 \lsun, more
characteristic of the sample of \citet{2008ApJ...678.1005P}. For
these models, we decreased the inner radius so that we could trace
the dust temperature to smaller radii. The dust temperature for the
L$_{int} = $ 0.7 \lsun\ protostar (Fig.~\ref{dustT:fig}) is lower
than 20 K everywhere outside about 400 AU. 
A detailed calculation of the chemistry preceding and during
the First Hydrostatic Core (FHSC) stage confirms this 
result \citep{2012arXiv1207.6693F}.
In contrast, the model with 150
\lsun\ has an extended region ($r < 8000$ AU)
with dust temperature above 20 K. If
the luminosities had always been less than or equal to the current
values, pure \cotwo\ ices should be prominent only in the high luminosity
sample, contrary to our observations.

We can also make a quantitative comparison of the current sample
and the sample from \citet{2008ApJ...678.1005P}. We
plot column densities of each \cotwo\ ice component
versus source luminosities in Fig.~\ref{lumino_correl}.  
We divide the sources for convenience into three groups:
the current, low-luminosity ($\lint < 0.7$) sample; the 
sources from \citet{2008ApJ...678.1005P} with intermediate 
luminosity ($ 0.7 < \lint < 100$ \lsun); and the high-luminosity
($\lint > \eten{3}$ \lsun) sources.
The sources in the
current study, with luminosity lower than 0.7 \lsun, have a similar
pure CO$_2$ column density range as the intermediate-luminosity sources. 
For example, the maximum pure CO$_2$ ice column
density detected toward sources with luminosity higher than 1.0
\lsun\ is $1.06 \times 10^{18}$ cm$^{-2}$, while the maximum pure
CO$_2$ ice column density is $6.8 \times 10^{17}$ cm$^{-2}$
in the source with internal luminosity 0.54 \lsun. 
The amount of pure \cotwo\ ice does not
depend on the current luminosity.

Table \ref{table-columnD} and Fig.~\ref{lumino_correl} show that the
H$_2$O-rich CO$_2$ ice component is dominant in all sources. Most of
the CO$_2$ ice exists as a mixture with water regardless of the
source luminosity. It takes more energy to segregate CO$_2$ ice and
H$_2$O ice, so H$_2$O-rich CO$_2$ ice is more stable than CO-CO$_2$
ice mixtures. In the low-luminosity sources, the mixture with CO exhibits
more scatter and higher values in some cases 
than in the high-luminosity sources. 
The sources with high luminosities may have evaporated much of the
CO from CO-CO$_2$ ice leaving pure CO$_2$ behind. 
The total \cotwo\ ice amount is also low in the high-luminosity 
sources compared the the spread seen in the low-luminosity sources. 
The high-luminosity sources 
could evaporate all the local \cotwo\ ices, leaving only the
ices in the extended cloud. The pure CO$_2$/total
CO$_2$ ratios have almost the same range of variation in the
low and intermediate-luminosity groups.
This result adds to the evidence for higher luminosities in the past
for the sources with current low luminosities.

\section{Discussion}\label{sec:Discuss}
\subsection{Chemical Models with Episodic Accretion}\label{sec:ChemModel}

One plausible reason for the presence of the pure CO$_2$ ice
component in low luminosity young stars is episodic accretion
\citep{1990AJ.....99..869K, 2005ApJ...633L.137V}. In this model, the
material falling in from the envelope is not continuously accreted
onto the star. Instead, it piles up on a disk and accretes onto a
star in a short-lived burst, when the disk becomes unstable. The
burst period can give enough heating to form  pure CO$_2$ ice in the
envelope. Once the pure CO$_2$ ice has formed, it will persist. A
series of ``freeze-thaw" cycles combined with irreversible chemical
processes can make distinct chemical signatures not possible in a
model with monotonic temperature behavior.

In addition to the pure \cotwo\ evidence, the total amount of
\cotwo\ ice, including that in mixed ices, can be explained by long
periods of low luminosity between episodic accretion bursts, as
predicted in an episodic accretion scenario. The source luminosity
steadily increases in a continuous accretion model as the mass of
the central star grows. Larger and larger fractions of the envelope
become too warm for freeze-out.  In contrast, the episodic accretion
model has multiple long periods of low luminosity
\citep{2010ApJ...710..470D}, punctuated by episodes of higher
luminosity. More ice can form during the low luminosity periods.

With chemical evolution modeling, we can relate the amount of
observed CO$_2$ ice to the accretion scenario. We use the
evolutionary chemo-dynamical model by \citet{2004ApJ...617..360L} to
test this idea. The model calculates the chemical evolution of a
core from the prestellar core to the embedded protostellar core
stage. At each time step, the density profile, the dust temperature,
the gas temperature, and the abundances are calculated
self-consistently.

The original chemical models can produce \cotwo\ ice only by
freezing of gas-phase \cotwo. We added a second route to \cotwo\ ice
production into the chemical network. Until recently there was no
known exact chemical pathway for CO$_2$ ice formation in cold
environments
\citep{1985A&A...152..130D,2001ApJ...555L..61R,2001MNRAS.324.1054R}.
Recently, \citet{2011ApJ...735...15G} suggested that CO ice turns
into CO$_2$ ice on cold interstellar dust grain surfaces in the
presence of OH radicals. Laboratory experiments support this idea
\citep{2010ApJ...712L.174O}. The process can occur at dust temperatures
as low as $\sim$ 10-12 K. In our study, we did not include the full
reaction of CO and OH radicals because the kinetics of \cotwo\ ice
formation on surfaces is an active field of research with no clear
answer so far. Instead we adopted a simple model. When the dust
temperature is below the CO ice freezing point, $(100-x)$\% of the
CO gas turns into CO ice, and $x$\% turns into \cotwo\ ice.

We ran two kinds of chemical models, those with and those without
the additional pathway to \cotwo\ ice from CO. For each kind of
chemical model, we ran two kinds of models for the luminosity
evolution, for a total of four kinds of models. The first luminosity
model assumes continuous accretion. We adopt the luminosity
evolution profile from \citet{2005ApJ...627..293Y} including the
First Hydrostatic Core stage (\citealp{1969MNRAS.145..271L,
1993ApJ...410..157B, 1998ApJ...495..346M, 2012arXiv1207.6693F}). 
Without the FHSC, the
luminosity evolution does not have a stage with luminosity lower
than 1.0 \lsun, so we use the model with the FHSC stage. In this
picture, the low-luminosity sources would be in the FHSC stage. The second
is an episodic accretion model from \citet{2010ApJ...710..470D}. In
this model, the episodic accretion is included in a simple,
idealized way. The model assumes no accretion from disk to star most
of the time, but steady infall from envelope to disk. When the disk
mass reaches 0.2 times the stellar mass, then an accretion burst
occurs with luminosity 100 times more than the usual value. We used
the model of a 1 \msun\ envelope from \citet{2010ApJ...710..470D}.
The luminosity evolutions of the continuous model and the episodic
accretion model are plotted in the top panels of Figs.~\ref{cont:evol}
and \ref{lum:evol}, respectively.
At early time steps, up to 20000 yrs, both the continuous accretion
model and the episodic accretion model evolve in the same way, going
through a FHSC (First Hydrostatic Core) stage. The luminosity
increases monotonically until it reaches 0.01 \lsun. After the FHSC
stage ends, the accretion luminosity increases rapidly, so there is
a luminosity gap between 0.01 \lsun\ and 1 \lsun\ in the continuous
accretion model and a gap between 0.01 \lsun\ and 0.1 \lsun\ in the
episodic accretion model.

At each time step, we calculate the dust temperature with DUSTY
\citep{1999astro.ph.10475I}, and the gas temperature with a gas
energetics code \citep{1997ApJ...489..122D,2004ApJ...614..252Y}
using the same modeling set up as \citet{2011ApJ...729...84K}. Then
the chemical evolution of 512 gas parcels is calculated as each
falls into the central region. We assumed the surface binding energy
of species onto bare SiO$_2$ dust grains. The chemical calculation
includes interactions between gas and dust and gas-phase reactions.
Then the abundance profile at each time step is calculated. To
compare models with ice observations, we calculated the total CO$_2$
ice column density at each time step.  We do not track the different
components of the \cotwo\ ice, as the chemical reactions between
them are not
well established, so we will compare the results to only the total
\cotwo\ ice.

To compare to the observations of gas phase CO, we use the
distributions of density, gas temperature, abundance, and velocity
at each time step to calculate the distribution of \cooo\ over its
energy levels and then simulate the observations by integrating
through the envelope and convolving the intensity with a beam
matched to the observations. Then we derive the ``observed" column
density of \cooo, using the same procedure as used for the
observations. The column density of gas phase CO is then obtained by
multiplying the \cooo\ column density by the isotope ratio (540;
\citealt{2004A&A...416..603J}). Because of the limited spatial
resolution of the observations and assumptions involved in
calculating the observed column density, this ``observed" column
density from the models will differ from the true column density,
but it provides the best comparison to observations.

Models were run with different values of $x$, the percentage of
freezing CO that goes into \cotwo\ ice. The best match to the
observations was found for $x = 10$, so we focus the rest of the
discussion on those models. First we discuss the model with a
constant accretion rate, followed by that with episodic accretion.

In the model with constant accretion (Fig.~\ref{cont:evol}) the
luminosity increases monotonically with time once the first
hydrostatic core forms (top panel). The other two panels in Figure
\ref{cont:evol} show the total column density in \cotwo\ ice and in
CO gas; they include the prestellar phase ($t < 0$), when the core
is assumed to pass through a series of Bonnor-Ebert spheres of
increasing central concentration.  Since evolution in this phase is
slow, the time axis is condensed. During this phase, the ``observed"
gas-phase CO increases by a factor of about 2 because the increasing
central density increases the excitation of the \cooo. The \cotwo\
ice column density increases as gas-phase \cotwo\ freezes out. These
processes accelerate at the later stages of highest central density,
but the freeze-out of CO begins to decrease the gas phase column
density toward the end of the prestellar evolution.  At $t = 0$, the
prestellar core evolution is ended by the formation of the FHSC. At
$t > 0$, the luminosity of the FHSC grows slowly, and some \cotwo\
ice evaporates, but it still dominates the gas phase CO column
density until the
true protostar forms at $t = 2\ee4$ yr. There is a slight increase
in the \cotwo\ ice toward the end of the FHSC phase owing to the
conversion of CO ice to \cotwo\ ice.  After $t = 2\ee4$ yr, the
luminosity increases as the stellar mass grows ($L \propto M$) and
the \cotwo\ gradually declines while the gas phase CO increases.
However, the \cotwo\ ice column density is still about twice the
gas-phase CO column density.

Fig.~\ref{lum:evol} shows the same graphs for the model of episodic
accretion. The evolution through the FHSC phase is identical to that
of the constant accretion model, but the differences begin with the
first episode of enhanced accretion. The dramatic increase in
luminosity causes a sharp increase in the gas phase CO, at the
expense of the \cotwo\ ice. After the accretion burst ends, the CO
begins to freeze again and the \cotwo\ ice builds up. This cycle is
repeated, with the balance of the C shifting between ice and gas
through each accretion cycle, but more ice forms in later quiescent
cycles because there is more time between bursts and the luminosity
between bursts is slightly lower as the steady state accretion rate
is lower. Also, some \cotwo\ ice persists in the cooler parts of the
envelope in each cycle. After the last accretion burst, the
luminosity declines slowly and the \cotwo\ ice increases at the
expense of the gas phase CO, but this process is limited by
processes that continually evaporate some CO. The evolution of
the ice and gas components is similar to that shown in 
\citet{2012ApJ...754L..18V}; differences can mostly be attributed to the
fact that our model follows the gas as it falls in, unlike
the single-point models in \citet{2012ApJ...754L..18V}.

\subsection{Comparison to Observations of Total \cotwo\ Ice}

In Fig.~\ref{Chemmodel}, we compare the model results and the
observational data. The green dots are the observed total \cotwo\
ice column density from both the current study and
\citet{2008ApJ...678.1005P}. The observed source luminosities range
from 0.1 \lsun\ to 90 \lsun. The continuous model does not have
luminosities higher than 2.4 \lsun, and the simple, bimodal,
episodic accretion model does not cover the luminosity range between
2.6 \lsun\ and 90 \lsun. Therefore, we compare the model and
observations only between 0.1 \lsun\ and 2.5 \lsun.

Both the original episodic accretion model and the continuous
accretion model, with the only path to \cotwo\ ice being freeze-out
of \cotwo\ gas, underestimate the observed column densities by an
order of magnitude. The detailed model of FHSC formation and 
evolution \citep{2012arXiv1207.6693F} confirms this result.
Matching the observations requires the extra
pathway in which CO ice is converted to \cotwo\ ice. Of the two
models with this pathway, the continuous model predicts only a small
range of \cotwo\ ice column densities. The episodic accretion model
with conversion of CO ice to \cotwo\ ice describes both the
luminosity spread and the spread in observed total \cotwo\ column
density reasonably well. 

The total column density of \cotwo\ ice at the end of the prestellar
core evolution is about $1 \times 10^{18}$ \cmn, is similar to column
densities seen toward background stars (e.g. \citealt{2009ApJ...695...94W}).
Consequently, the accumulation of \cotwo\ ice in the quiescent cloud, 
before the formation of the core would roughly double the column density 
in our models, which do not explicitly include this evolution.
Differences from cloud to cloud and from YSO to YSO in the amount
of this absorption from the quiescent cloud could explain some of 
the scatter, allowing the continuous accretion model to fit the data 
better. However, much of the scatter lies in the regime of 
$N(\cotwo)$ above the values usually seen toward background stars
\citep{2009ApJ...695...94W}.

\subsection{Comparison to Observations of \cooo\ Gas}

Another check on the models is provided by the residual amount of
\cooo\ gas. Leaving out the two sources with the highest CO column
density, the average \cooo\ gas column density is $9.18 \times
10^{14}$ cm$^{-2}$. We round off to  $\sim$ 10$^{15}$ cm$^{-2}$ as a
residual \cooo\ gas column density, which is plotted as a horizontal
line in Fig.~\ref{cogas:evol}. The four models are plotted in four
different panels for comparison to the data. After a certain amount
of time, the \cooo\ gas predicted by the constant accretion models
converges to a roughly constant value, $N \sim 6 \times 10^{14}$
cm$^{-2}$, lower than the mean value. In the episodic accretion
model, \cooo\ gas evaporates after each episodic accretion burst,
but $N \sim$ 10$^{15}$ cm$^{-2}$ most of the time, as observed.
Of course, \cooo\ emission can also come from the larger cloud.
An example may be source 182, with the largest \cooo\ column density;
this source is in the Ophiuchus cloud, which has large column densities
even along lines of sight that do not intercept YSOs.

The different predictions arise from the different abundance
profiles. The abundance profiles at time step 60000 yr, the last
time step in the continuous accretion model, are plotted in
Fig.~\ref{abundance:60000}, along with those from the episodic
accretion models, which are in a high luminosity state at this time.
The high luminosity has heated the envelope, and thus raised the gas
phase CO abundance out to much larger radii, compared to the
continuous accretion model.

The column density versus luminosity graphs (Fig.~\ref{cogas:lum})
compare the models to observations. Seven sources are scattered
around the predictions of the model with episodic accretion and CO
to \cotwo\ ice conversion. Two sources lie well above and two lie
well below the predictions. The source with the least amount of \cooo\ gas is
CB130-1. That source was best fit with a model in which 80\% of the
CO ice turns into \cotwo\ ice  \citep{2011ApJ...729...84K}, which is
much higher than used here. Most source are better matched if 10\%
of CO ice turns to \cotwo\ ice.

The current models provide a proof of concept for episodic accretion
models to explain the observed \cotwo\ ice and gas-phase CO
abundances, but they can be improved by future work. The episodic
accretion included in this study is a simple and idealized model
with extreme luminosity variations and does not include the absorption
from the larger cloud. It does not cover the luminosity
range between 2.3 \lsun\ and 90 \lsun. Fig.~\ref{Chemmodel} shows
that there is a luminosity gap between 2.3 \lsun\ and 90 \lsun\ in
the episodic accretion scenario, while the observed sources do have
luminosities in the range between 3 \lsun\ and 100 \lsun. Using more
realistic models of the accretion history and a range of envelope
masses (\citealt{2011ApJ...736...53O, 2012ApJ...747...52D})
will allow more detailed comparison to the observations.

\subsection{Is Episodic Accretion a Unique Explanation?}

In this paper we have shown that a scenario of episodic accretion as
modeled by \citet{2010ApJ...710..470D} can explain the presence of a
component of pure \cotwo\ ice in the observed 15.2 \micron\
absorption feature, the total column  density of all CO$_2$ ice, and
the column density of gas-phase C$^{18}$O in a  sample of
low-luminosity protostars.  Coupled with the ability of such a
scenario to also explain the observed protostellar luminosity
distribution \citep{2009ApJ...703..922V, 2010ApJ...710..470D,
2012ApJ...747...52D} and with the  existence of other, indirect
  observational 
evidence for variable accretion (Dunham \& Vorobyov 2012 and
references therein), there seems to be strong  evidence in favor of
the existence of episodic accretion.

However, models predicting other mass accretion and luminosity histories 
have also been proposed.  
As first noted by \citet{1990AJ.....99..869K}, 
a model in which accretion rates peak early and then decline 
through the remainder of the protostellar phase
could potentially match the observed protostellar luminosity distribution.
Models of this type have been advanced by a number of authors
(e.g., \citealt{1999ApJ...513L..57A,2011ApJ...736...53O,2012ApJ...747...22H}).
\citet{2012ApJ...747...52D} recently argued that such a scenario by itself is 
insufficient to match observed protostellar luminosities, although their 
conclusions were based only on the specific set of models considered in their 
study.  
A model with an initial peak in the luminosity followed by a long period of 
decline might reproduce the observations if the initial peak were high enough 
to produce sufficient pure CO$_2$ ice and the time spent at lower luminosities 
were long enough to convert sufficient CO ice to CO$_2$ ice.  Full chemical 
modeling of such an accretion scenario is required to fully evaluate how well 
it could reproduce our observations, but this is beyond the scope of our 
present study.  While we acknowledge that such a scenario 
could possibly explain our ice and gas-phase observations, we 
consider it somewhat unlikely given the existence of sources like 
IRAM04191$+$1522 (source 003) that clearly show double-peaked features 
indicative of pure CO$_2$ ice. These are still deeply embedded, 
Class 0 sources,
too young to have already spent sufficient time in the low-luminosity, 
declining accretion phase in typical models of this type 
(e.g., \citealt{1999ApJ...513L..57A}).

Other accretion scenarios may also explain observed protostellar
luminosities. \citet{2011ApJ...736...53O} presented analytic derivations 
of the protostellar luminosity distribution for different models and 
concluded that models that tend toward a constant accretion time rather than 
constant accretion rate produce a greater spread in luminosities and are in 
better agreement with observations. \citet{2012ApJ...747...22H} presented 
simulations including protostellar feedback, both protostellar
radiation and outflows, and argued that such simulations can match
the means, medians, and standard deviations of the observed protostellar 
luminosity distribution without episodic accretion.
However, they did not compare the full observed and modeled 
luminosity distributions.
The accretion rates in \citet{2012ApJ...747...22H} 
and a subset of the models in \citet{2011ApJ...736...53O},
the subset they called the ``tapered'' models, feature 
an early peak, followed by a slow decline in the accretion rate 
throughout the remainder of the embedded phase.  
As discussed above, such models could possibly reproduce our 
observations, although such models will also be challenged to 
predict the existence of very young, deeply embedded, 
very low luminosity protostars with significant pure CO$_2$ ice.

Detailed calculations of the chemistry for specific evolutionary
histories will be needed to test these other models. One possible
problem with models that feature one period of high luminosity
followed by a long period of low luminosity is that CO ice will
form over the pure \cotwo\ ice, rendering it less visible.

Ultimately, our results do not prove that episodic accretion occurs.  They do, 
however, demonstrate that the past accretion rates must have been higher 
than the current rates for about half of our targets, and they do show 
that a scenario of episodic accretion is fully consistent with our 
observations.

\section{Conclusion}\label{sec:conclusion}
We detected the pure \cotwo\ ice double-peaked feature from 6 out of
19 low luminosity sources  with L $\leq 1$ \lsun. Fits to the
absorption profile provide strong evidence for pure \cotwo\
ice in 3 more sources, for a total of 9 (47\%). 
The minimum required temperature to form pure
CO$_2$ ice is 20 K, which is not found in any substantial part of the
envelope at the current luminosities of these sources.  Detection of
pure CO$_2$ ice strongly indicates higher luminosities in the
past. During  the time of higher luminosity, the pure CO$_2$ ice forms
and persists into the current low luminosity stage.
The existence of pure \cotwo\ ice toward such low-luminosity sources
is very strong evidence of a phase of higher luminosity in the
past, consistent with models of episodic accretion. It also
presents a challenge to any alternative model for explaining the
luminosity spread seen toward YSOs.

In addition to the luminosity spread of sources
(\citealt{2010ApJ...710..470D}, Dunham and Vorobyov 2011), the total
CO$_2$ ice amount and the residual gas phase CO are best explained
by episodic accretion with conversion of CO to CO$_2$ ice. If 10\%
of CO goes into \cotwo\ ice when the gas phase CO freezes,
observations of most sources are reasonably well matched.
There is a substantial spread of values of both total \cotwo\ ice
and gas phase \cooo\ in the observations, which can be explained in 
an episodic chemistry. However, variations in the ice formation in 
the larger cloud can also contribute to this scatter. Determinations
of other ice components and extinction in the surrounding cloud can
help to distinguish these effects.

This material is based upon work supported by the National
Aeronautics and Space Administration under RSA 137730 issued by the
Jet Propulsion Laboratory. Jeong-Eun Lee was supported by Basic
Science Research Program through the National Research Foundation of
Korea(NRF) funded by the Ministry of Education, Science and
Technology (No. 2012-0002330). Hyo Jeong Kim and Neal J Evans
acknowledge support from NSF grant AST-0607793.

\clearpage

\begin{deluxetable}{cccrrcr}
\tabletypesize{\scriptsize}
\tablecaption{The low luminosity sources observed with \emph{Spitzer} IRS} 
\tablewidth{0pt} 
\tablehead{
\colhead{Source Number\tablenotemark{a}} & \colhead{\emph{Spitzer}
Source Name} & \colhead{c2d Core Region} &  \colhead{RA} &
\colhead{DEC} & \colhead{Internal luminosity\tablenotemark{b}} &
\colhead{PID}
}
\startdata
 & & & (J2000) & (J2000) & (L$_{\odot}$) & \\

\hline \\
003 & SSTc2d J042200.41+153021.2 & IRAM04191$+$1522 & 04:22:00.41 & $+$15:30:21.2 &   0.08 & 20604\\
005 & SSTc2d J044112.65+254635.4 &    TMC1          & 04:41:12.65 & $+$25:46:35.4 &   0.36 & 50295\\
016 & SSTc2d J124539.96-552522.4 & DC302.1+7.4      & 12:45:39.96 & $-$55:25:22.4 &   0.42 & 50295\\
017 & SSTc2d J130736.89-770009.7 & DC303.8-14.2     & 13:07:36.89 & $-$77:00:09.7 &   0.46 & 50295\\
021 & SSTc2d J165719.63-160923.4 &    CB68    &   16:57:19.63 & $-$16:09:23.4 &   0.54 & 50295\\
023 & SSTc2d J171122.18-272602.0 &    B59     &   17:11:22.18 & $-$27:26:02.0 &   0.42 & 50295\\
024 & SSTc2d J181616.39-023237.7 &    CB130-1 &   18:16:16.39 & $-$02:32:37.7 &   0.15 & 50295\\
038 & SSTc2d J212407.58+495908.9 &    L1014   &   21:24:07.58 & $+$49:59:08.9 &   0.09 & 20604\\
043 & SSTc2d J222959.52+751403.1 &    L1251   &   22:29:59.52 & $+$75:14:03.1 &   0.21 & 50295\\
060 & SSTc2d J032637.46+301528.1 &    Perseus &   03:26:37.46 & $+$30:15:28.1 &   0.69 & 50295\\
063 & SSTc2d J032738.26+301358.8 &    Perseus &   03:27:38.26 & $+$30:13:58.8 &   0.20 & 50295\\
068 & SSTc2d J032845.29+310542.0 &    Perseus &   03:28:45.29 & $+$31:05:42.0 &   0.26 & 50295\\
078 & SSTc2d J032923.47+313329.5 &    Perseus &   03:29:23.47 & $+$31:33:29.5 &   0.20 & 50295\\
090 & SSTc2d J033229.18+310240.9 &    Perseus &   03:32:29.18 & $+$31:02:40.9 &   0.20 & 50295\\
092 & SSTc2d J033314.38+310710.9 &    Perseus &   03:33:14.38 & $+$31:07:10.9 &   0.14 & 50295\\
109 & SSTc2d J034421.36+315932.6 &    Perseus &   03:44:21.36 & $+$31:59:32.6 &   0.11 & 50295\\
124 & SSTc2d J125342.86-771511.5 & Chamaeleon II &     12:53:42.86 & $-$77:15:11.5 &   0.14 & 50295\\
161 & SSTc2d J160115.55-415235.4 &    Lupus   &   16:01:15.55 & $-$41:52:35.4 &   0.08 & 50295\\
182 & SSTc2d J162705.23-243629.5 & Ophiuchius &   16:27:05.23 & $-$24:36:29.5 &   0.15 & 50295\\
\enddata
\tablenotetext{a}{The source number comes from
  \citet{2008ApJS..179..249D}.}
\tablenotetext{b}{The source luminosity comes from the
  relation between 70 \micron~  flux and internal source luminosity
  \citep{2008ApJS..179..249D}.}
\label{table-source}
\end{deluxetable}


\begin{deluxetable}{crrrrrrr}
\tabletypesize{\scriptsize} \tablecolumns{8} \tablewidth{0pc}
\tablecaption{Ice column densities of the CO$_2$ component}
\tablehead{ \colhead{Source Number} & \colhead{Pure CO$_2$} &
\colhead{CO:CO$_2$=100:4} & \colhead{CO:CO$_2$=100:26} &
\colhead{CO:CO$_2$=100:70} & \colhead{H$_2$O:CO$_2$=100:14} &
\colhead{Total CO$_2$} &
\colhead{$\chi^2$}\\
}
\startdata
003\tbnm{a} &  1.717 $\pm$  0.130 &  0.063 $\pm$  0.005 &  0.305 $\pm$  0.023 &  1.336 $\pm$  0.106 &  5.949 $\pm$  0.417 &  9.370 $\pm$  0.680 &  0.034\\
005 &  0.297 $\pm$  0.014 &  0.000 $\pm$  0.001 &  0.801 $\pm$  0.031 &  1.992 $\pm$  0.063 &  8.108 $\pm$  0.245 & 11.198 $\pm$  0.353 &  0.920\\
016 &  3.507 $\pm$  0.414 &  9.235 $\pm$  0.968 &  0.000 $\pm$  0.002 &  7.687 $\pm$  0.883 & 22.288 $\pm$  2.233 & 42.717 $\pm$  4.500 &  0.980\\
017\tbnm{b} &  1.864 $\pm$  0.103 &  0.086 $\pm$  0.008 &  0.071 $\pm$  0.005 &  8.186 $\pm$  0.444 & 25.543 $\pm$  1.348 & 35.751 $\pm$  1.908 &  4.124\\
021\tbnm{a} &  6.837 $\pm$  0.497 &  0.000 $\pm$  0.002 &  0.010 $\pm$  0.012 &  8.072 $\pm$  0.610 & 24.355 $\pm$  1.731 & 39.274 $\pm$  2.853 &  1.604\\
023 &  0.696 $\pm$  0.032 &  0.000 $\pm$  0.001 &  0.464 $\pm$  0.019 &  6.027 $\pm$  0.085 & 11.985 $\pm$  0.123 & 19.172 $\pm$  0.260 &  8.218\\
024\tbnm{a} &  2.751 $\pm$  0.150 &  0.256 $\pm$  0.015 &  0.343 $\pm$  0.026 &  3.387 $\pm$  0.181 & 11.762 $\pm$  0.594 & 18.500 $\pm$  0.966 &  0.847\\
038 &  1.460 $\pm$  0.087 &  0.623 $\pm$  0.057 &  0.222 $\pm$  0.011 &  2.550 $\pm$  0.120 & 14.490 $\pm$  0.440 & 19.345 $\pm$  0.715 &  1.399\\
043\tbnm{b} &  0.535 $\pm$  0.016 &  0.000 $\pm$  0.001 &  0.000 $\pm$  0.001 &  2.050 $\pm$  0.035 &  4.673 $\pm$  0.058 &  7.258 $\pm$  0.109 &  2.090\\
060 &  1.506 $\pm$  0.061 &  0.000 $\pm$  0.001 &  0.001 $\pm$  0.010 &  7.813 $\pm$  0.355 & 19.302 $\pm$  0.785 & 28.622 $\pm$  1.212 &  3.518\\
063 &  0.401 $\pm$  0.023 &  0.000 $\pm$  0.001 &  0.500 $\pm$  0.023 &  2.410 $\pm$  0.063 &  6.124 $\pm$  0.066 &  9.435 $\pm$  0.175 &  1.135\\
068\tbnm{a} &  3.778 $\pm$  0.133 &  0.000 $\pm$  0.001 &  0.005 $\pm$  0.009 &  7.724 $\pm$  0.262 & 14.364 $\pm$  0.434 & 25.870 $\pm$  0.838 &  2.166\\
078\tbnm{a} &  3.070 $\pm$  0.129 &  0.000 $\pm$  0.001 &  0.123 $\pm$  0.073 &  6.054 $\pm$  0.321 & 13.141 $\pm$  0.568 & 22.389 $\pm$  1.091 &  2.011\\
090 &  0.161 $\pm$  0.003 &  0.000 $\pm$  0.001 &  0.260 $\pm$  0.024 &  0.777 $\pm$  0.051 &  3.842 $\pm$  0.050 &  5.041 $\pm$  0.128 &  3.480\\
092 &  0.347 $\pm$  0.019 &  0.000 $\pm$  0.001 &  0.689 $\pm$  0.123 & 12.807 $\pm$  0.704 & 20.304 $\pm$  1.036 & 34.148 $\pm$  1.882 &  1.948\\
109 &  0.017 $\pm$  0.014 &  0.000 $\pm$  0.001 &  0.782 $\pm$  0.048 &  1.252 $\pm$  0.061 &  3.860 $\pm$  0.058 &  5.910 $\pm$  0.182 &  3.767\\
124\tbnm{a} &  0.723 $\pm$  0.032 &  0.000 $\pm$  0.001 &  0.000 $\pm$  0.001 &  1.036 $\pm$  0.053 &  3.548 $\pm$  0.074 &  5.308 $\pm$  0.159 &  1.432\\
161\tbnm{b} &  3.540 $\pm$  0.261 &  0.000 $\pm$  0.001 &  0.000 $\pm$  0.001 &  4.357 $\pm$  0.271 & 16.695 $\pm$  1.003 & 24.591 $\pm$  1.536 &  0.802\\
182 &  0.000 $\pm$  0.001 &  0.000 $\pm$  0.001 &  1.329 $\pm$  0.033 &  0.868 $\pm$  0.034 &  3.378 $\pm$  0.071 &  5.575 $\pm$  0.139 &  3.848\\
\enddata
\tablecomments{All column densities are in 10$^{17}$ cm$^{-2}$.}
\tbnt{a}{Clearly double-peaked feature.}
\tbnt{b}{High signal-to-noise column density of pure \cotwo\ feature
and $\chi^2$ improves by more than 20\% when pure \cotwo\ is included.}
\label{table-columnD}
\end{deluxetable}


\begin{deluxetable}{crrrrrrrr}
\tablecolumns{9}
\tablewidth{0pc}
\tablecaption{
The \cooo ~(\jj{2}1) observation of low luminosity embedded sources}
\tablehead{
\colhead{Source Number} &
\colhead{rms}&
\colhead{$\int T^*_A dv$}&
\colhead{$v_{LSR}$} &
\colhead{$\Delta v$} &
\colhead{$T^*_A$} &
\colhead{Column Density} &
\colhead{Reference} \\
 \colhead{} &
 \colhead{(K)} &
 \colhead{(K km s$^{-1}$)} &
 \colhead{(km s$^{-1}$)} &
 \colhead{(km s$^{-1}$)} &
 \colhead{(K)} &
 \colhead{(cm$^{-2}$)} &
 \colhead{}}

\startdata
003 &  \nodata& 1.68  &  6.38 & 0.76 & 2.06 & 1.31e+15 & (1)\\
005 &   0.138 & 1.31  &  5.24 & 0.70 & 1.77 & 1.02e+15 &\\
021 &   0.144 & 1.25  &  5.03 & 0.52 & 2.29 & 9.69e+14 &\\
023 &   0.146 & 3.13  &  3.45 & 1.03 & 2.86 & 2.44e+15 &\\
024 &   0.078 & 0.48  &  7.61 & 0.71 & 0.64 & 3.74e+14 & (2) \\
043 &   0.216 & 1.20  & -4.22 & 0.82 & 1.36 & 9.31e+14 &\\
060 &   0.140 & 0.58  &  5.11 & 0.52 & 1.05 & 4.52e+14 &\\
063 &   0.135 & 1.22  &  5.20 & 2.59 & 0.44 & 9.49e+14 &\\
068 &   0.141 & 0.99  &  8.34 & 1.41 & 0.66 & 7.72e+14 &\\
109 &   0.138 & 1.71  &  8.96 & 0.84 & 1.90 & 1.33e+15 &\\
161 &   0.172 & 1.39  &  4.02 & 0.77 & 1.70 & 1.08e+15 &\\
182 &   0.199 & 4.69  &  4.72 & 1.20 & 3.66 & 3.65e+15 &\\
\enddata
\tablecomments{Reference: (1) Chen et al (2012, in prep), (2) \citet{2011ApJ...729...84K}}
\label{table:molecularline}
\end{deluxetable}

\clearpage

\clearpage

\begin{figure}[t]
\includegraphics{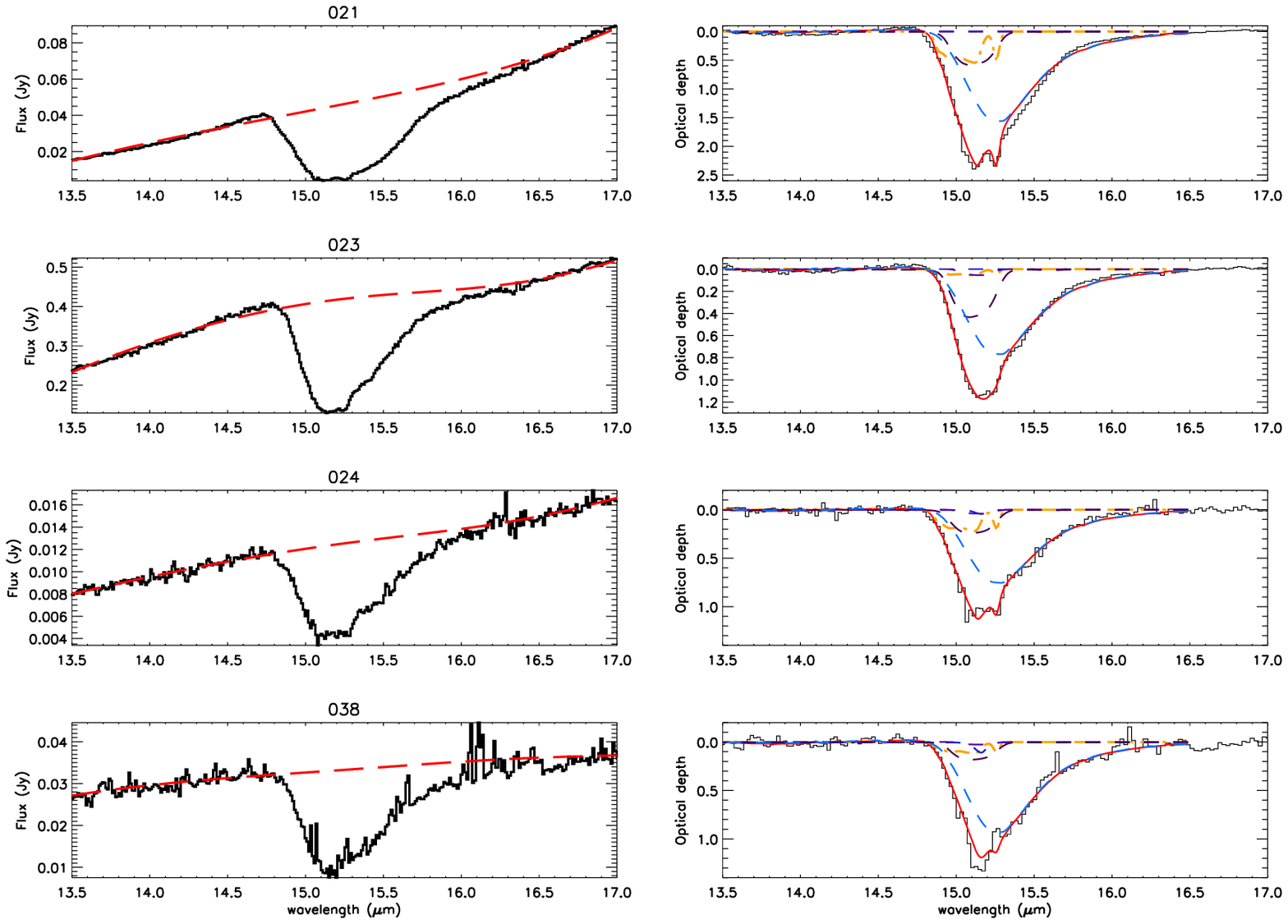}
\caption{\label{figure2:spectrum} Continued}
\end{figure}
\clearpage

\begin{figure}[t]
\includegraphics{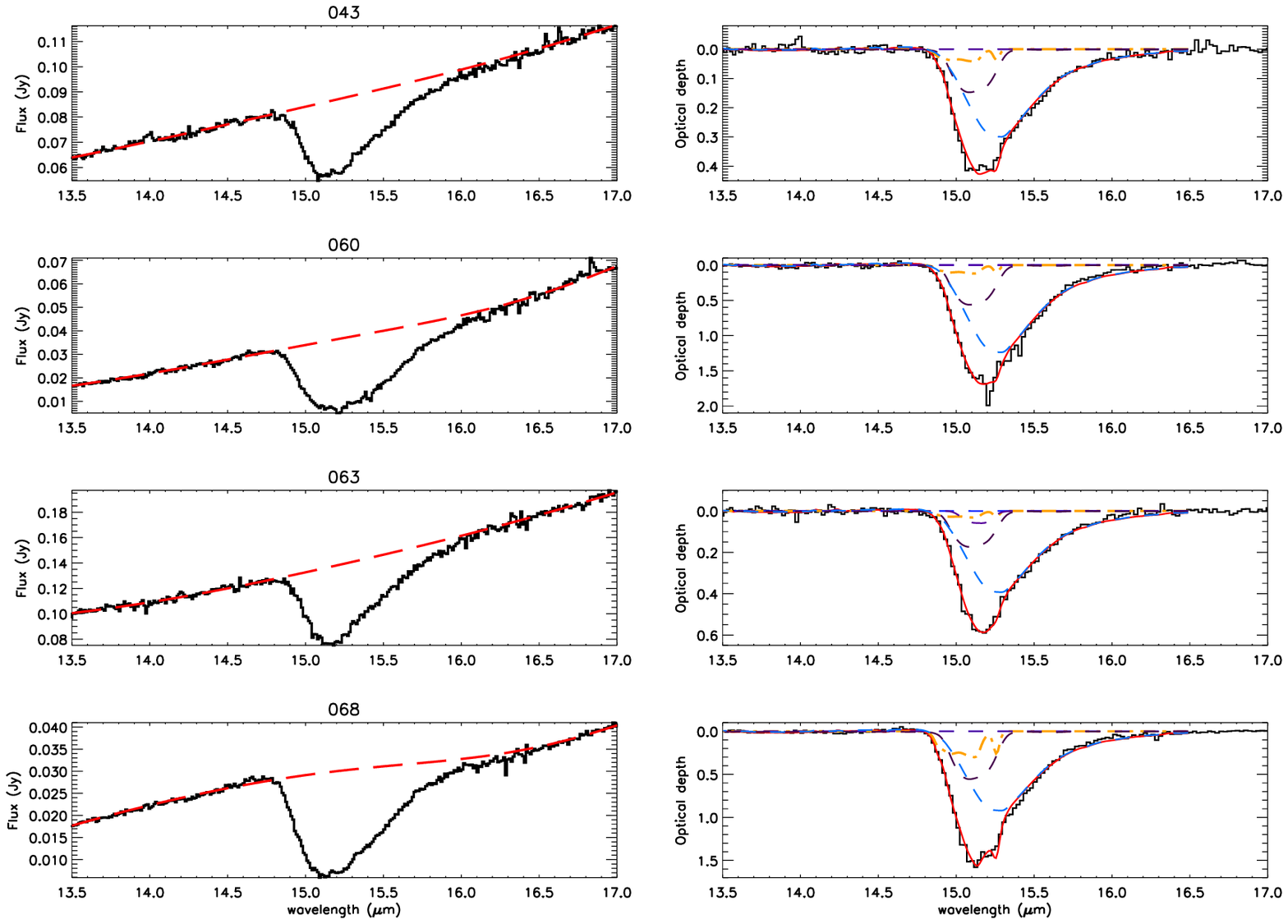}
\caption{\label{figure3:spectrum}Continued}
\end{figure}
\clearpage

\begin{figure}[t]
\includegraphics{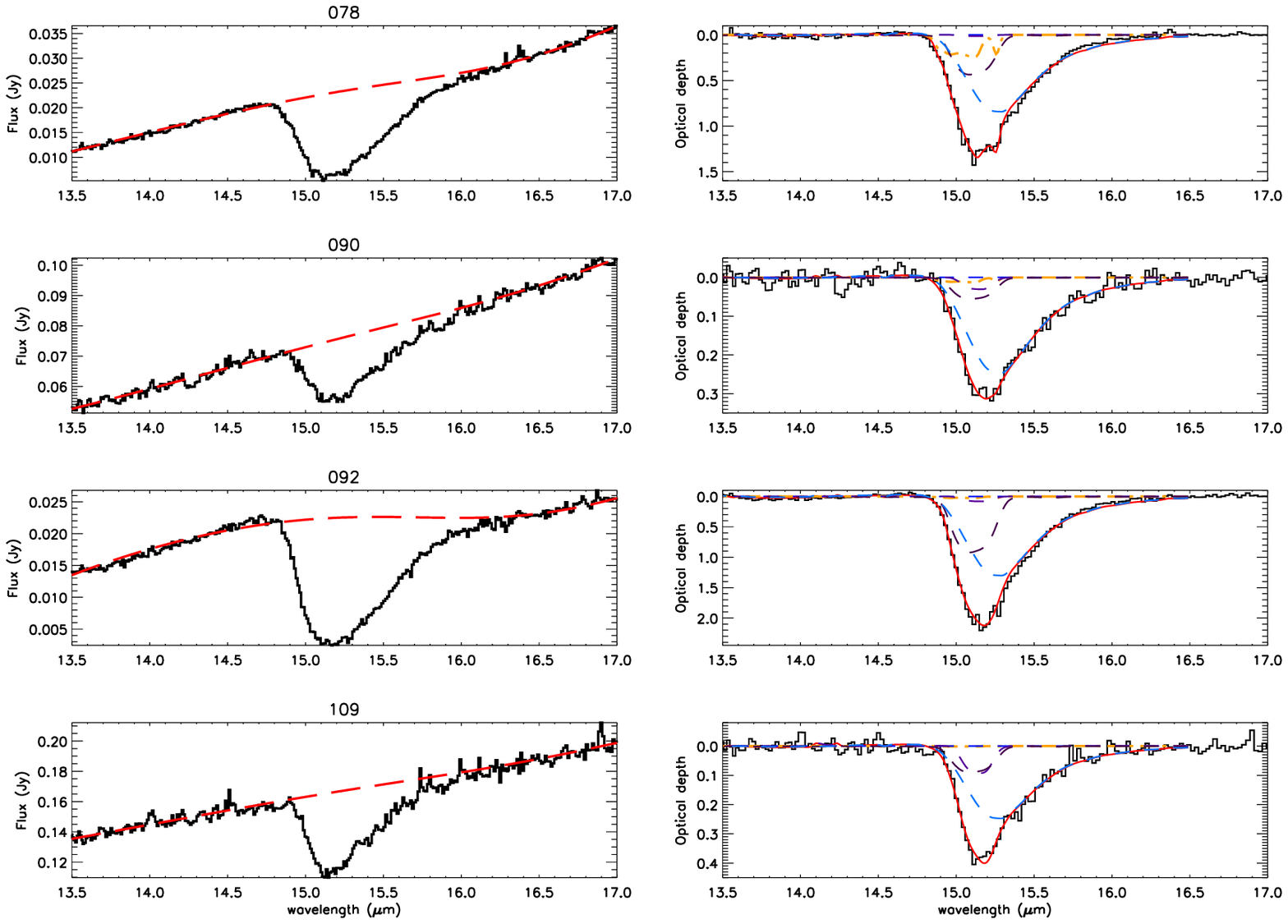}
\caption{\label{figure4:spectrum}Continued}
\end{figure}
\clearpage

\begin{figure}[t]
\includegraphics{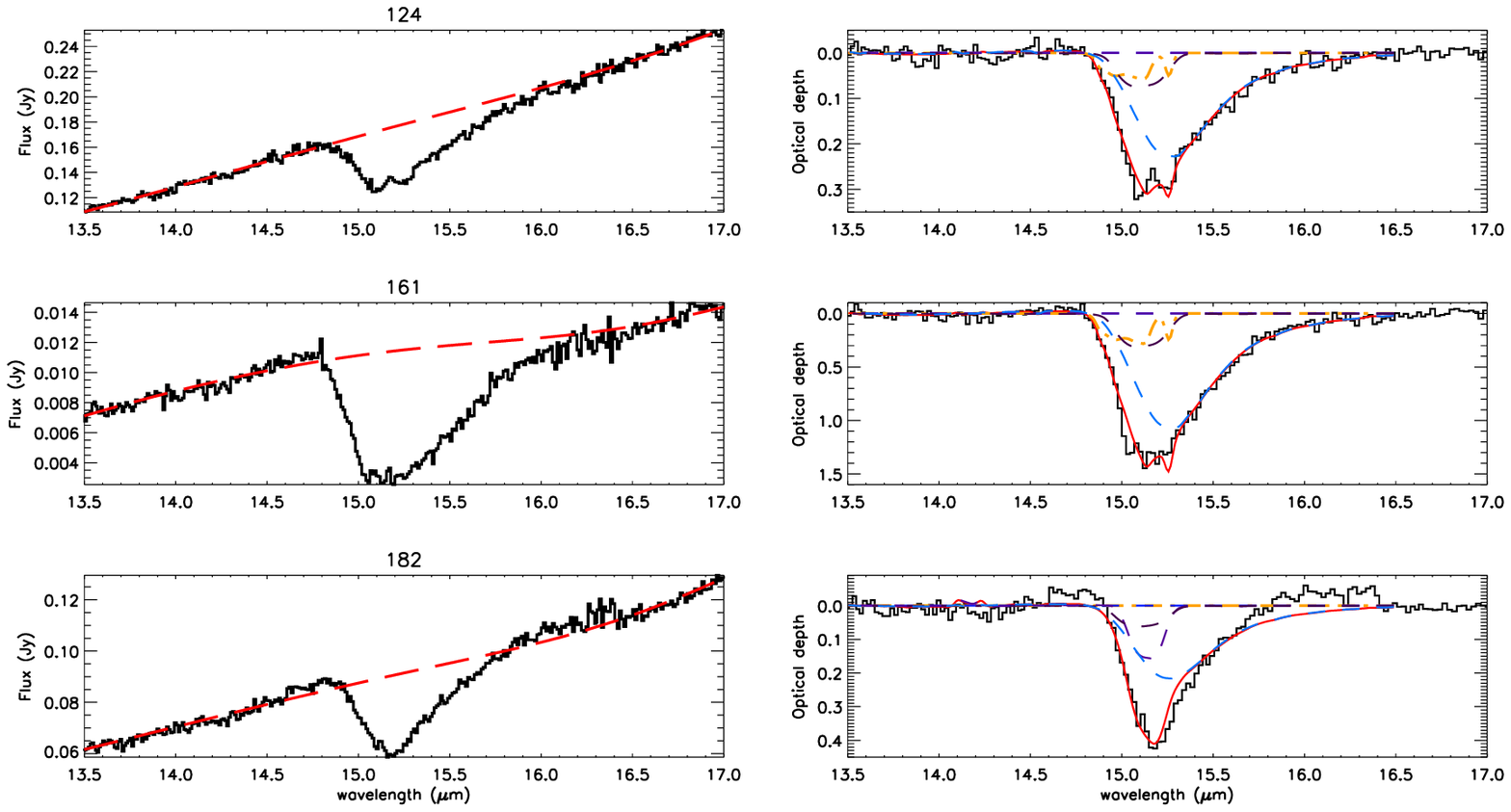}
\caption{\label{Ophiuchius:spectrum}Continued}
\end{figure}
\clearpage

\begin{figure}[t]
\includegraphics{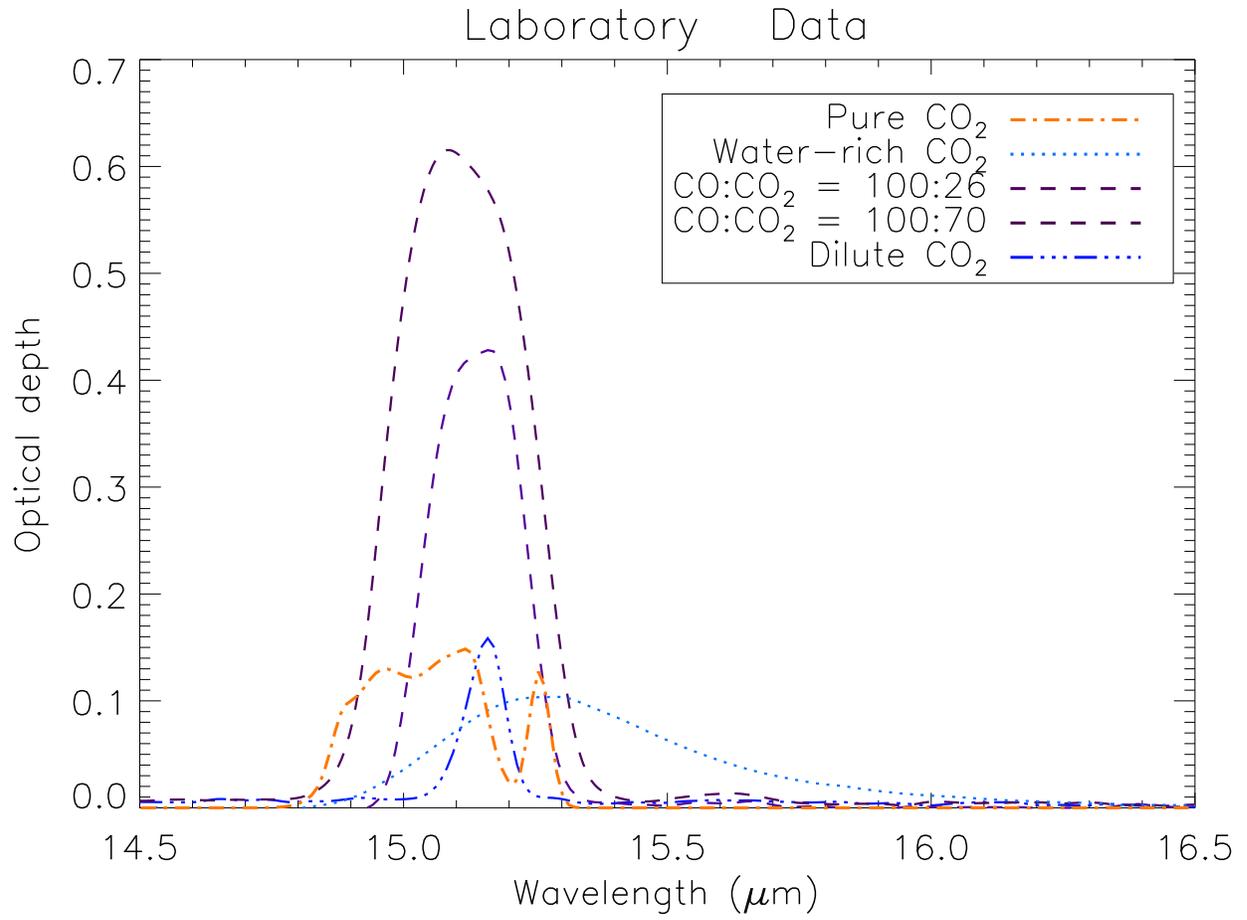}
\caption{\label{labo:dat} A plot of laboratory data. The yellow line
  is a pure  CO$_2$ ice, the light blue line is a water-rich  CO$_2$ ice,
  the blue line is a dilute  CO$_2$ ice, and purple lines are CO:CO$_2$
  ice mixtures. Only the pure CO$_2$ ice component shows a double
  peak feature. A baseline is subtracted from a water-rich CO$_2$ ice component.}
\end{figure}
\clearpage

\begin{figure}[t]
\includegraphics[scale=0.8]{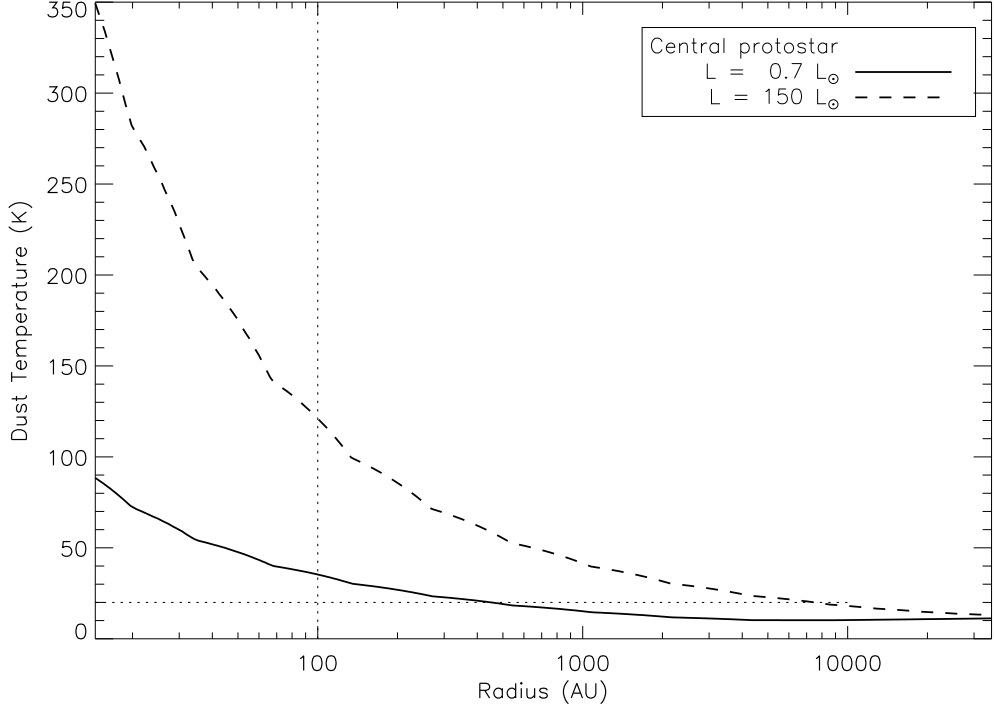}
\caption{\label{dustT:fig}
Dust temperatures as a function of radius for two different luminosities,
characteristic of the current sample ($L = 0.7$ \lsun) and the previous sample
($L = 150$ \lsun). The inner radius in both models is 10 AU, the same as
was used in the chemical models used in this paper. The envelope inner
radii of embedded protostars are about 100 AU. The horizontal dashed line
indicates a dust temperature of 20 K, the minimum required for producing
pure \cotwo\ ice.
}
\end{figure}
\clearpage

\begin{figure}[t]
\includegraphics{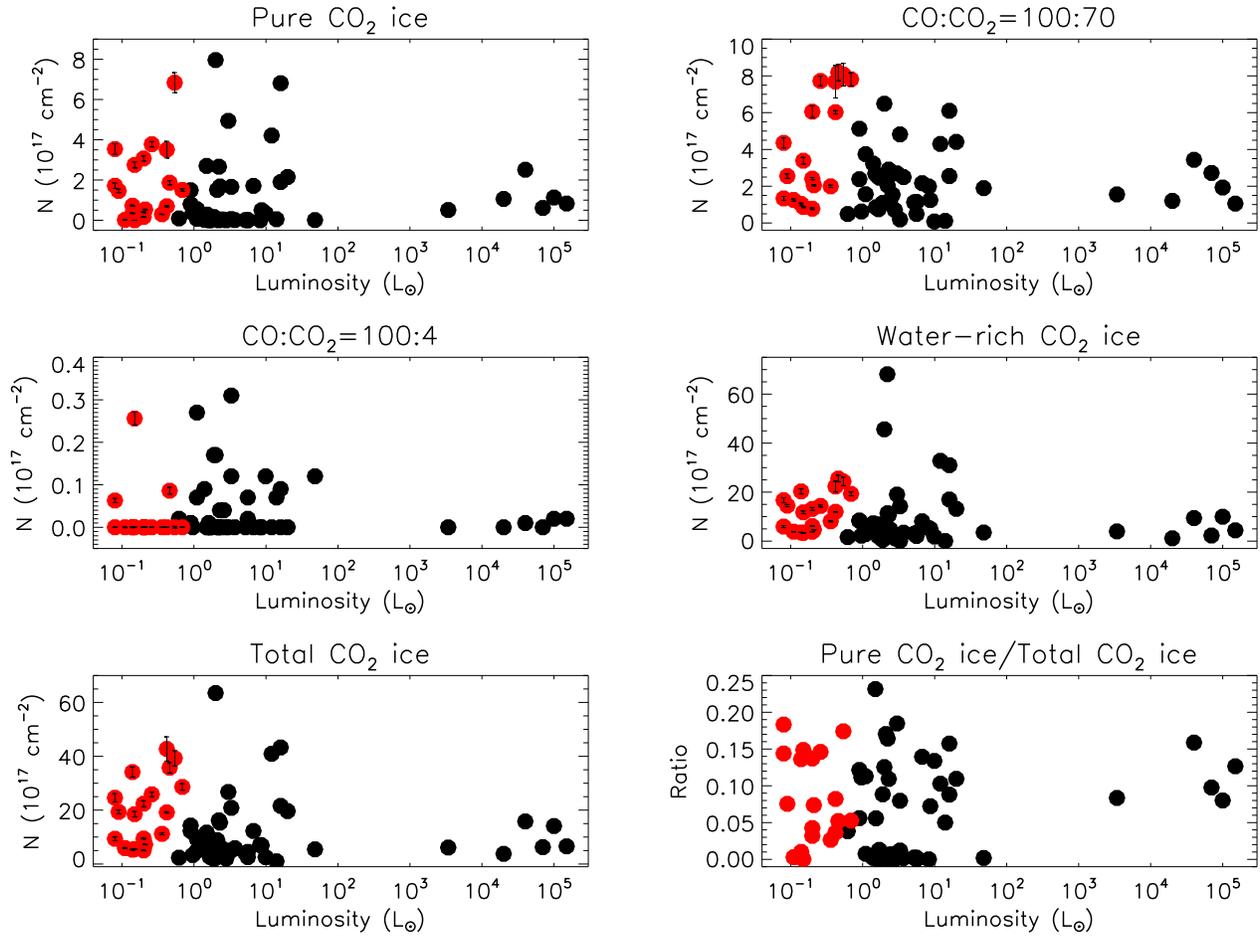}
\caption{\label{lumino_correl}
The column densities of the different \cotwo\ ice components and the total
plotted versus source luminosities. The red points are from the current study;
the black points are from \citet{2008ApJ...678.1005P} }
\end{figure}
\clearpage

\begin{figure}[t]
\includegraphics{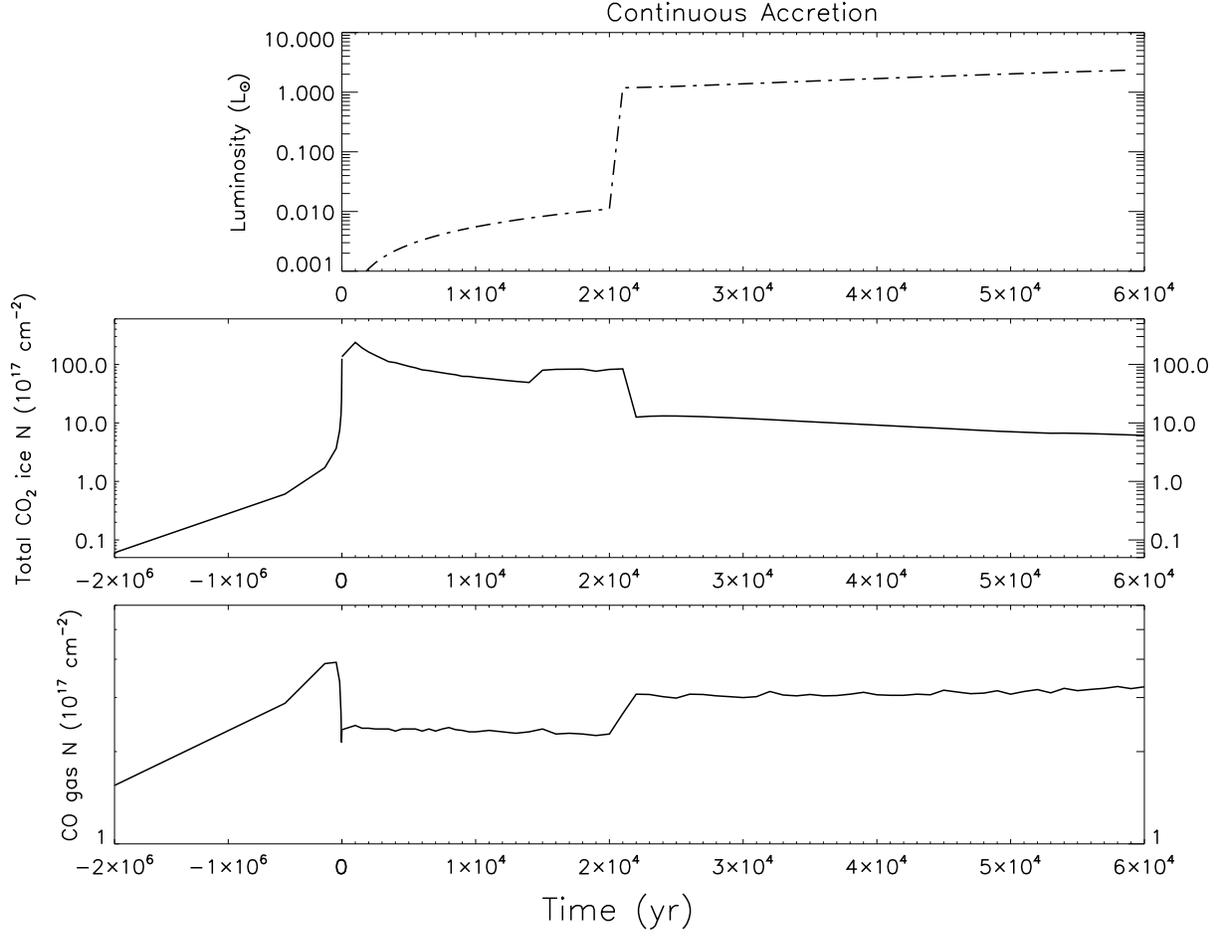}
\caption{\label{cont:evol} Top panel: The luminosity evolution of the
continuous accretion model.
Middle panel: The \cotwo\ ice column density evolution in the constant
accretion model
with 10\% CO ice to \cotwo\ ice conversion.
Bottom panel: The gas phase CO column density evolution in the constant
accretion model
with 10\% CO ice to \cotwo\ ice conversion.
The chemical model includes a long prestellar core stage and a FHSC stage.
The time axis is compressed when $t<0$  since evolution is slow during
that time.
}
\end{figure}
\clearpage

\begin{figure}[t]
\includegraphics{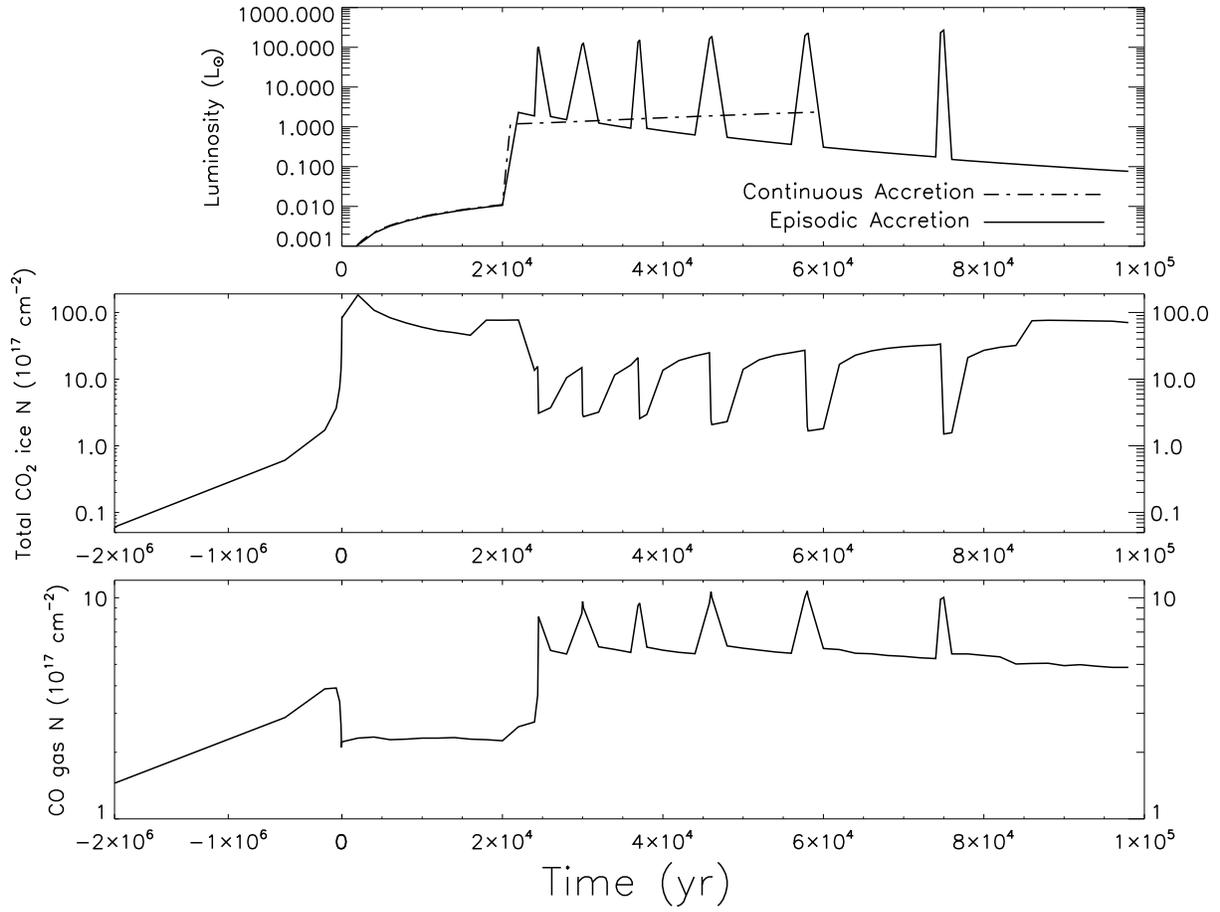}
\caption{\label{lum:evol} Top panel: The luminosity evolution of the
episodic accretion model. The dash-dot line is the luminosity of the
continuous accretion model for comparsion.
Second panel: The total \cotwo\ ice column density evolution
in the episodic accretion model with
10\% CO ice to \cotwo\ ice conversion.
Third panel: The gas phase CO column density evolution
in the chemical model including the episodic accretion model with
10\% CO ice to \cotwo\ ice conversion.
}
\end{figure}
\clearpage

\begin{figure}[t]
\includegraphics{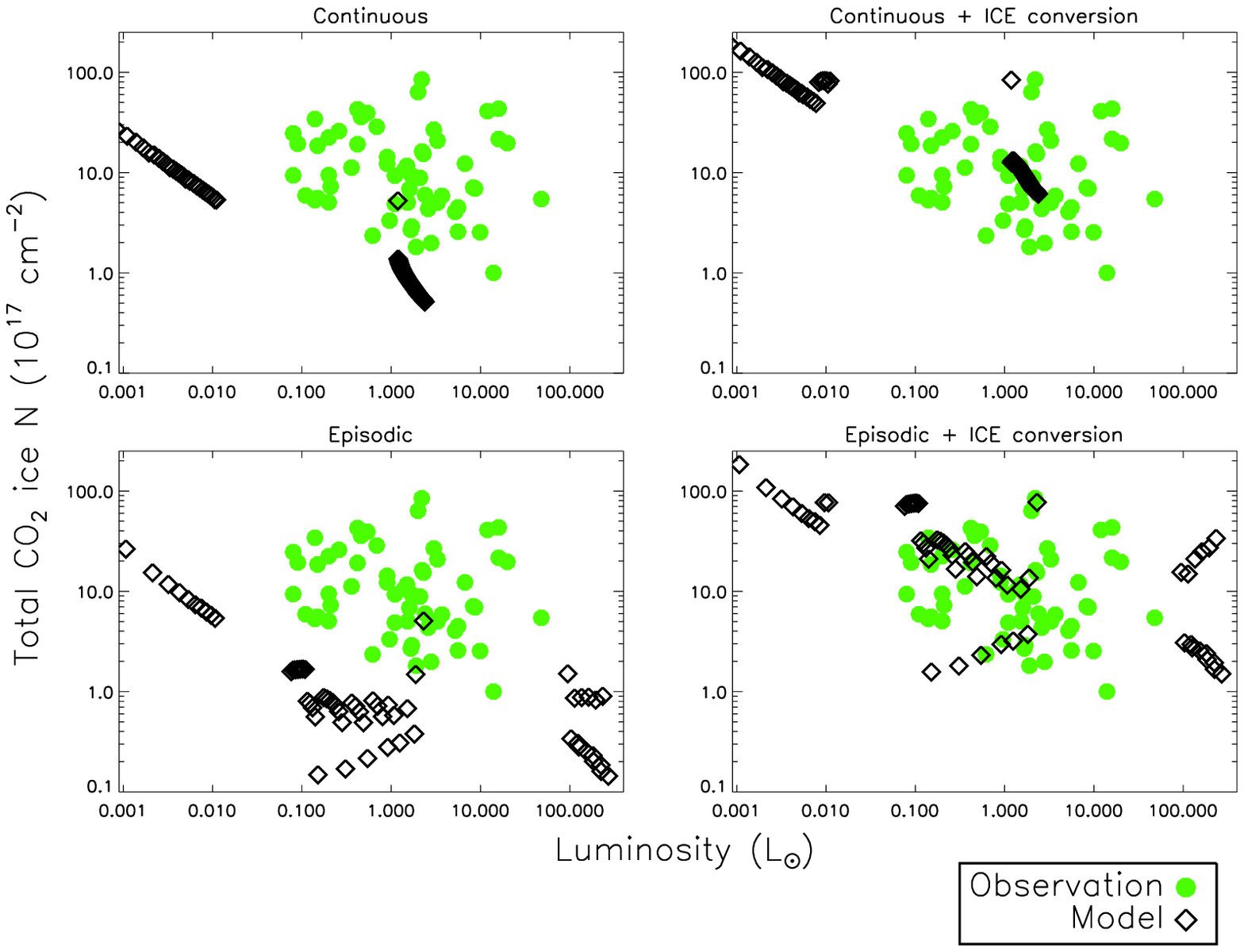}
\caption{\label{Chemmodel}
The column densities of \cotwo\ ice (total) from observations and
each of the four luminosity-chemical model combinations.
The upper panels
have continuous accretion while the lower panels have episodic
accretion. The left panels have standard chemical models while the
right panels include the extra pathway for \cotwo\ ice formation.
The observed column densities are plotted as green dots, while the
model predictions are plotted as black diamonds.
}
\end{figure}
\clearpage

\begin{figure}[t]
\includegraphics{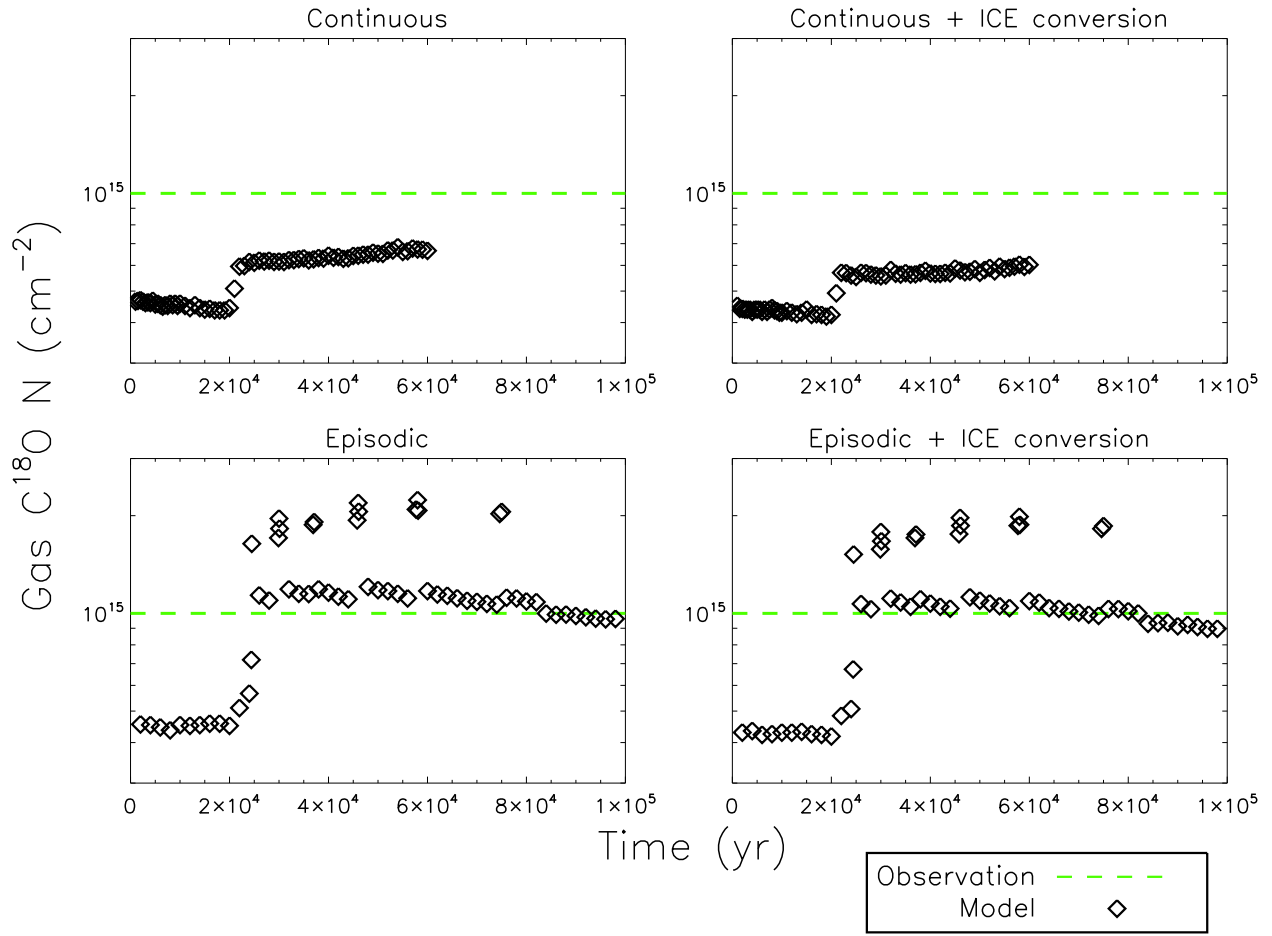}
\caption{\label{cogas:evol}
The gaseous \cooo\ column density
from each of the four luminosity-chemical models
is plotted versus time.
The upper panels
have continuous accretion while the lower panels have episodic
accretion. The left panels have standard chemical models while the
right panels include the extra pathway for \cotwo\ ice formation.
The dashed line is the typical observed column density of \cooo\ gas.
}
\end{figure}
\clearpage

\begin{figure}[t]
\includegraphics{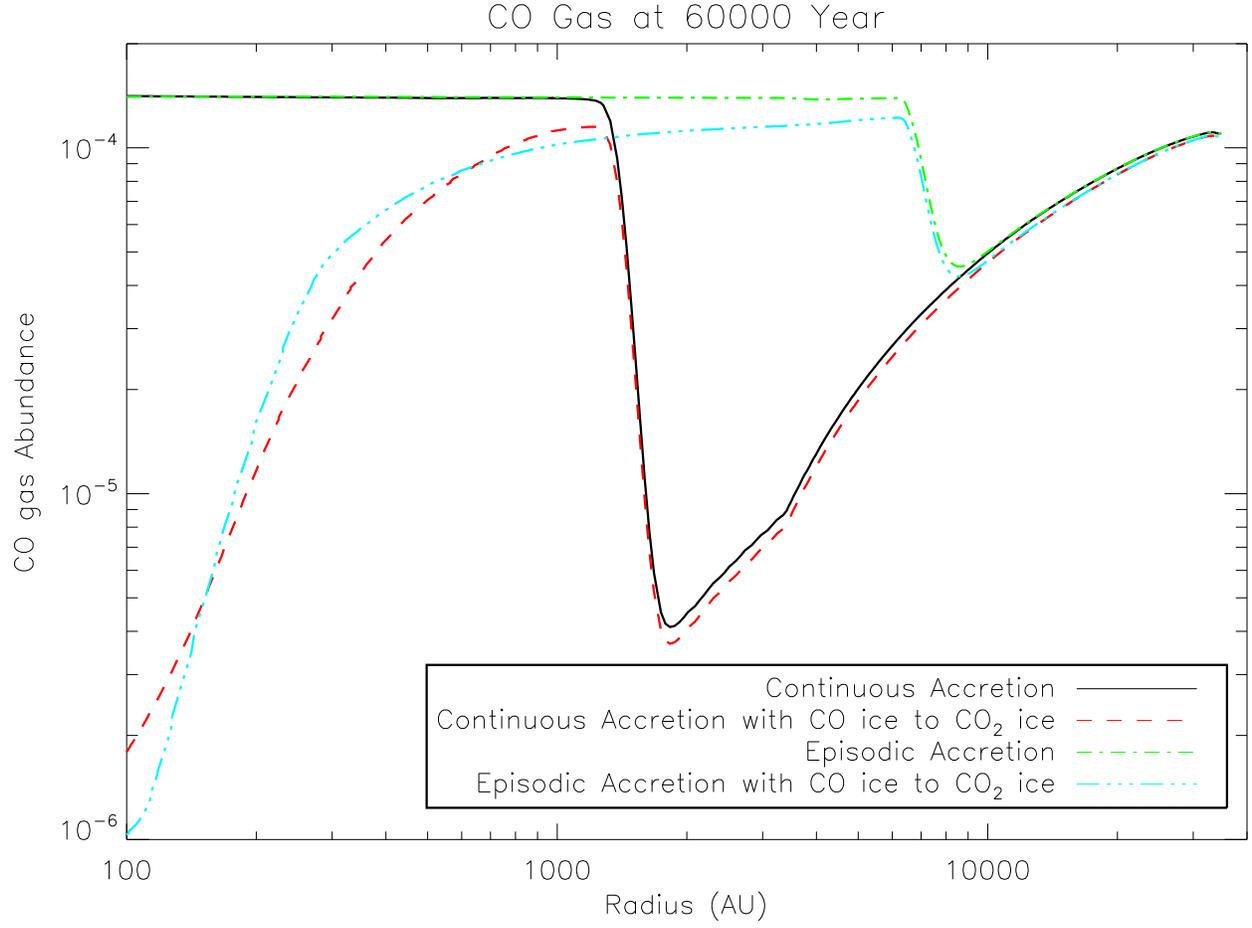}
\caption{\label{abundance:60000} The gaseous CO abundance profile at
  60000 yr from each of the four luminosity-chemical models.
  The time step is the last time step of the continuous
  accretion model. At this time, the episodic accretion model is in
  a high luminosity state, and the region of high gas phase CO abundance
extends to much larger radii than in the continuous accretion model.
However, the gas phase CO is still depleted in the inner regions due to
conversion into \cotwo\ ice.
}
\end{figure}
\clearpage

\begin{figure}[t]
\includegraphics{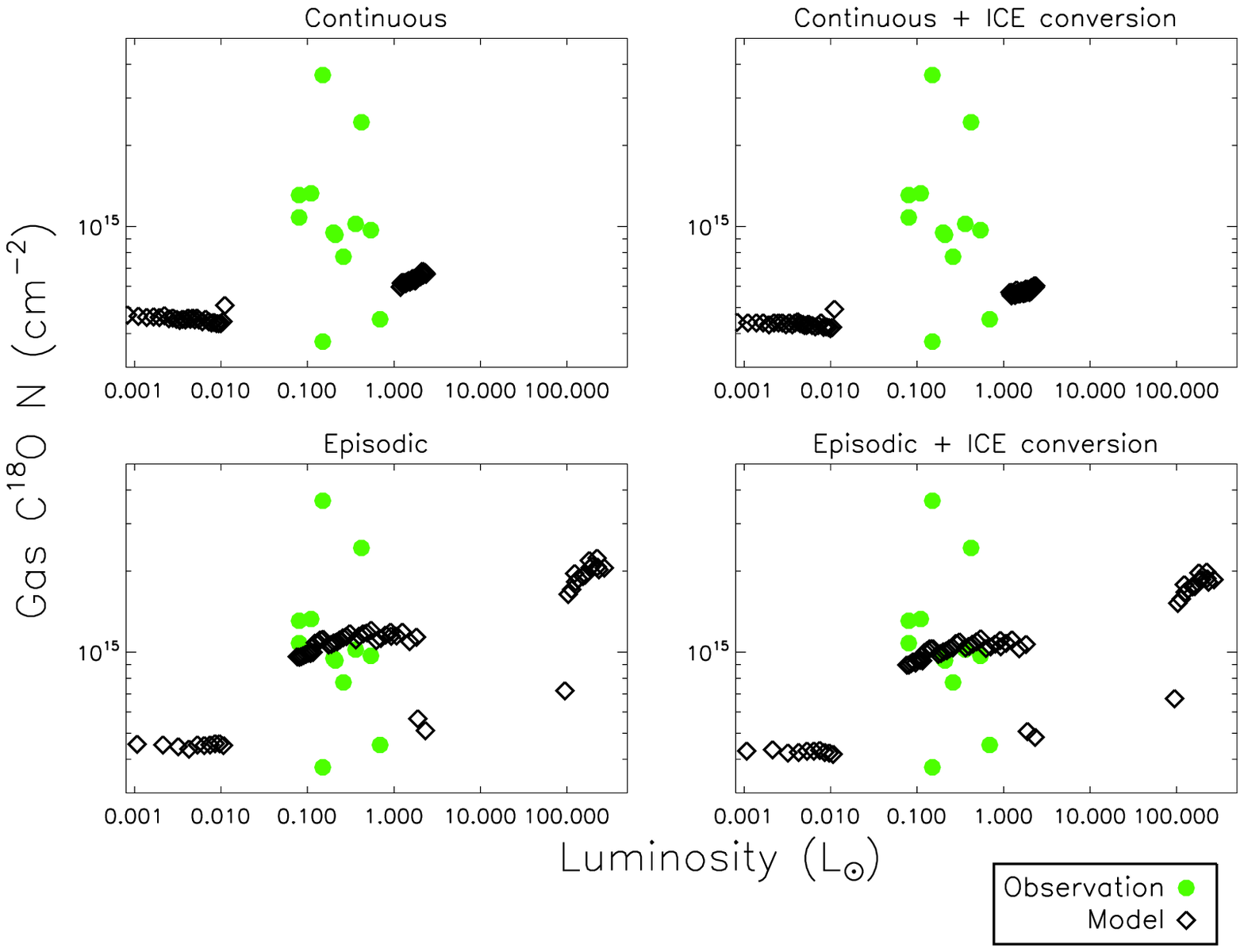}
\caption{\label{cogas:lum}
The column densities of \cooo\ gas from observations and
each of the four luminosity-chemical model combinations.  The upper panels
have continuous accretion while the lower panels have episodic
accretion. The left panels have standard chemical models while the
right panels include the extra pathway for \cotwo\ ice formation.
The observed column densities are plotted as green dots, while the
model predictions are plotted as black diamonds.
}
\end{figure}
\clearpage

\end{document}